\documentclass[fleqn,usenatbib]{mnras}

\usepackage{newtxtext,newtxmath}
\usepackage{comment}

\usepackage[T1]{fontenc}

\DeclareRobustCommand{\VAN}[3]{#2}
\let\VANthebibliography\thebibliography
\def\thebibliography{\DeclareRobustCommand{\VAN}[3]{##3}\VANthebibliography}

\usepackage{graphicx}	
\graphicspath{ {./images/} }
\usepackage{amsmath}	
\usepackage{pifont}
\usepackage{mathtools}
\usepackage{placeins}
\usepackage{float}
\usepackage{bm}
\usepackage{multirow}
    
\usepackage{nccmath}
\usepackage{xcolor}
\usepackage{soul}

\usepackage{cleveref}
\crefname{figure}{Fig.}{Figs.}
\crefname{table}{Table}{Tables}

\definecolor{orange}{rgb}{1,0.5,0}
\definecolor{sred}{rgb}{.5,0,0}

\title[Protostellar disks in their natural habitat]{Protostellar disks in their natural habitat - the formation of protostars and their accretion disks in the turbulent and magnetized interstellar medium}

\author[Mayer et al.]{%
Alexander C. Mayer$^{1,2}$\thanks{E-mail: amayer@mpa-garching.mpg.de}, Thorsten Naab$^{1}$, Paola Caselli$^{3,2}$, Alexei V. Ivlev$^{3,2}$, Tommaso Grassi$^{3,2}$, Oliver Zier$^{4}$, \newauthor Rüdiger Pakmor$^{1}$, Stefanie Walch$^5$, Volker Springel$^{1}$
\vspace{0.1cm}\\%
$^{1}$Max-Planck-Institut für Astrophysik, Karl-Schwarzschild-Straße 1, 85741 Garching, Germany\\%
$^{2}$Excellence Cluster ORIGINS, Boltzmannstraße 2, 85748 Garching, Germany\\%
$^{3}$Max-Planck-Institut f\"ur Extraterrestrische Physik, Giessenbachstr. 1, 85748 Garching, Germany \\
$^{4}$Center for Astrophysics | Harvard \& Smithsonian, 60 Garden St, Cambridge, MA 02138, USA\\
$^{5}$I. Physikalisches Institut, Universität zu Köln, Zülpicher Str. 77, 50937 Köln, Germany \\
}

\date{Accepted XXX. Received YYY; in original form ZZZ}

\pubyear{2025}

\begin{document}
 \label{firstpage}
\pagerange{\pageref{firstpage}--\pageref{lastpage}}
\maketitle

\begin{abstract}
We present simulations of the supernova-driven turbulent interstellar medium (ISM) in a simulation domain of volume $(256\,{\rm pc})^3$ within which we resolve the formation of protostellar accretion disks and their stellar cores to spatial scales of $\sim 10^{-4}$ au, using the moving-mesh code {\small AREPO}. We perform simulations with no magnetic fields, ideal magnetohydrodynamics (MHD) and ambipolar diffusion, and compare the resulting first Larson cores and their associated structures, including the accretion disks, their location within the larger-scale structure and the streamers connecting these. We find that disks of sizes $10-100\,{\rm au}$ form early in the simulations without magnetic fields, while there are no disks larger than 10 au with ideal MHD. Ambipolar diffusion causes large disks to form in a subset of cases (two out of six cores), and generally reduces the strength of outflows, which are seen to play a central role. When they are able to carry away significant angular momentum, they prevent the formation of a rotationally supported disk. Magnetic fields strengths grow from $0.1 - 1$ mG in the protostellar core to more than 10 G in the first Larson core in all simulations with ideal MHD. The rotationally supported disks which form can have rotation speeds $> 1$ km s$^{-1}$ even out to further than 100 au from the centre, become gravitationally unstable and form complex spiral substructures with Toomre $Q < 1$. We conclude that the impact of magnetic fields and non-ideal MHD on the formation of protostellar disks is substantial in realistic formation scenarios from the turbulent ISM.
\end{abstract}


\section{Introduction}
Star formation is a highly multi-scale process, where turbulence in the interstellar medium cascades down from galactic scales to molecular clouds and cloud cores, thereby ultimately impacting the formation of protostars and their disks \citep[see e.g. the reviews by][]{maclow2004starformation,mckee2007starformation,chevance2023clouds,hacar2023ism,pineda2023bubbles}. As such, processes occurring at the kpc-scale might affect the stellar initial mass function (IMF) \citep{brucy2020turbulence}, which in turn influences the driving of turbulence set mostly by feedback processes like supernova explosions \citep{joung2006ism,kim2015ism,walch2015silcc,guszejnow2022feedback}. Therefore, a large range of scales needs to be modelled in simulations that aim to capture the full process of protostar formation. 

In contrast, the most well-studied scenario for the formation of protostars and their surrounding accretion disks is the collapse of {\em isolated} turbulent molecular cloud cores, considering the impact of radiation, magnetic fields, non-ideal magnetohydrodynamics (MHD) effects, or even all simultaneously \citep[e.g.,][]{bate2003starcluster,commercon2010radiative,bate2014collapse,zhao2016dust,grudic2021starforge,wurster2021nonidealimpactsingle}. 
The resulting overall picture of star formation from the collapse of a molecular cloud core \citep[based on the fundamental work of][]{larson1969}, has been well-established numerically at this point, with a suggested progression in approximately four stages. There is first an approximately isothermal collapse phase where support is almost entirely provided by magnetic forces. In the classical strong-field picture, a flattened `pseudo-disk' supported by magnetic forces is formed \citep{Galli&Shu1993I,Galli&Shu1993II} due to the anisotropic infall caused by the magnetic field. However, the formation of a prominent pseudo-disk requires the magnetic field to be dynamically dominant over turbulent motions in the core, which stands in contrast to more recent measurements of the magnetic field strength in giant molecular clouds \citep{crutcher2010fields,pattle2023bfield}. With a further increase in density in the central region, it eventually becomes optically thick, allowing a pressure-supported structure to form, called the first (Larson) core \citep{hennebelle2008magnetic,xu2021formationI,xu2021formationII}. As more mass is accreted, the temperature increases further, and at temperatures of $2000\,{\rm K}$ dissociation of H$_2$ commences. Because this reaction is endothermal, the pressure rises much slower as a function of density and the effective adiabatic index approaches $\gamma \approx 1.1$, much softer than in the first hydrostatic core. The protostar is born as the equation of state becomes dominated by atomic hydrogen, leading to a rise in the adiabatic index ($\gamma = \frac{5}{3}$)  and the formation of a pressure-supported structure -- known as the second (Larson) core -- within the first core at roughly the size of a solar radius \citep{tomida2013rmhd,ahmad2023protostar,ahmad2025birth}. 

The inclusion of magnetic fields is motivated both by observational evidence of strong magnetization in star-forming regions that affects the dynamics at the filament and core scale \citep{pattle2023bfield} and because the magnetic field is also thought to be dynamically important for limiting disk sizes by removing angular momentum during the collapse of the protostellar core \citep{wurster_li2018magnetic,zhao2020review}. If flux-freezing of the magnetic field in the collapsing gas is assumed (\textit{ideal MHD}),  magnetic fields have been predicted by some models to completely stop the formation of protostellar accretion disks in what is called the `magnetic braking catastrophe', as seen in early ideal MHD calculations \citep{allen2003braking,galli2006braking}. However, by now it has been consistently shown in numerical studies that the change in the magnetic field evolution brought about by non-ideal MHD effects in the form of Ohmic dissipation, ambipolar diffusion and the Hall effect can allow the formation of disks after all \citep[e.g.,][]{zhao2016dust,vaytet2018protostellar,wurster2021nonidealimpactsingle}. These three effects are macrophysical descriptions of the fact that, unlike what is assumed in ideal MHD, not all particles respond instantaneously to changes in the magnetic field. Instead, the resistivity is finite, the fraction of neutral particles is non-zero, and the inertia differs between different charge carriers \citep[see e.g.,][]{wardle1999conductivity,pandey_wardle2008,zhao2018grains}. 

In present-day cold molecular cloud cores, the cosmic-ray ionization rate is at locally inferred to be low at $\zeta_i < 10^{-16}$ s$^{-1}$ \citep{redaelli2021l1544,pineda2024ngc1333} \citep[with a `canonical' rate of $\zeta_i = 10^{-17}$ s$^{-1}$, ][]{padovani2009cirate}. \cite{pineda2024ngc1333} find a typical electron fraction of $10^{-6.5}$, consistent with theoretical considerations inferring low ionization degrees in molecular clouds \citep{Shu1992book,Tielens2005book,mckee2007starformation}. Hence, molecular cloud cores are weakly ionized,  indicating that the influence of non-ideal MHD is substantial. However, there have also been numerous studies concluding that the `magnetic braking catastrophe' can be prevented -- even in ideal MHD -- by considering more complex collapse scenarios of isolated cloud cores without including any larger-scale environment. The simplest extension which appears to allow disk formation is an initial misalignment of magnetic field and rotation axis \citep{hennebelle2009misalignment,joos2012misalignment}, which can naturally occur by including turbulence in the initial conditions \citep{santos-lima2012turbulence,myers2013fragmentation,seifried2013turbulence}. However, there is no consensus on the relative importance of non-ideal MHD effects and turbulence when both are considered simultaneously \citep{seifried2013turbulence,wurster2019catastrophe,wurster2020turbI}. In contrast to these studies, \cite{bate2018diversity} finds good agreement of the statistics of disk sizes when simulating the collapse of a low-mass molecular cloud with no magnetic field at all, arguing that this may indicate that, in reality, non-ideal MHD effects are sufficiently strong to cause a close similarity to purely hydrodynamical simulations. 

Observations over the last years have revealed that protostellar accretion disks are usually not exclusively fed through a relatively homogeneous spherical or, under the influence of magnetic fields, sheet-like accretion. Instead, there exist small filaments of material denser than their surroundings, known as \textit{streamers} \citep[e.g.,][]{pineda2020streamer,valdivia-mena2022streamer,hsieh2023streamer,flores2023streamer,podio2024streamer,choudhurry2025streamer}. These streamers deposit material onto disks at specific spots, thus departing strongly from a simple picture of relatively symmetric accretion. They also cause heating via accretion shocks and increase disk masses locally, which can in turn trigger gravitational instabilities. 

Many previous multi-scale studies of star formation under the influence of turbulence either do not resolve the actual formation of protostars \citep[often using large sink particles, e.g.,][]{grudic2021starforge}, or do not include the large scale ISM where the turbulence is actually driven and cascades down from to strongly influence disk and protostar formation  \citep[e.g.,][]{walch2012imf,bate2025imf}. 
In this study, we aim to improve on this by computing simulations of stellar core formation that start from the scale of the supernova-driven turbulent ISM and are large enough that the expansion of supernova explosions can be properly followed, using pure hydrodynamics, ideal MHD, and ambipolar diffusion. Our simulation domains are (256 pc)$^3$ in size, which is more than 10 orders of magnitude larger than the solar radius (the typical initial size of a second Larson core). We only use sink particles to identify regions as `zoom-in targets', which we then re-simulate to actually resolve the collapse of the protostellar core down to the formation of the second Larson core, with a minimum cell size smaller than $10^{-3}$ au.

Our simulations generally follow the approaches of \cite{kueffmeier2017zoom} and \cite{seifried2017silcczoom}. While we utilize a simpler ISM model for our initial turbulent driving, we employ a larger simulation volume, have a higher maximum resolution (competitive even compared to isolated collapse studies, and resolving up to the stellar core), and we compare the cases of ideal and non-ideal MHD. More recently, \cite{yang2025zoom} have performed a zoom-in on a single protostellar disk, starting from a box stirred with driven turbulence, only employing ideal MHD and with a maximum resolution of 1 au. Both \cite{kueffmeier2017zoom} and \cite{yang2025zoom} focused on the longer-term evolution of disks, which we do not consider. 

We note that there have also been studies focusing on the statistics of disks formed in turbulent and magnetized massive star-forming clumps \citep{lebreully2021magnetic,lebreully2024population}, including the impact of ambipolar diffusion. Other works considered lower-mass clumps, while studying the impact of non-ideal MHD processes in more detail \citep{wurster2019catastrophe}. Both lines of work found that the inclusion of non-ideal MHD effects influences disk structure and size. \cite{ahmad2025birth}, on the other hand, focused on resolving the formation and early evolution of the stellar core, both in ideal and non-ideal MHD (including ambipolar and Ohmic diffusion) in the collapse of an initially turbulent cloud core.

This work is structured as follows. In Section~\ref{sec:methods}, we describe the simulation code and the physical processes that we use in our model, while the actual setup and details of the simulations are found in Section~\ref{sec:simulations}. We then analyze the prestellar cores that we have chosen as zoom-in regions in Section~\ref{sec:cores}. We move on to the main results, which concern the properties of the disks and hydrostatic cores formed within those zoom-in regions, in Sections~\ref{sec:results} and \ref{sec:results_specific}. After a discussion of the limitations and possible extensions of our model in Section~\ref{sec:discussion} we summarize our conclusions in Section~\ref{sec:conclusions}.
\section{Methods}
\begin{figure*}
    \centering
    \includegraphics[width=1\linewidth]{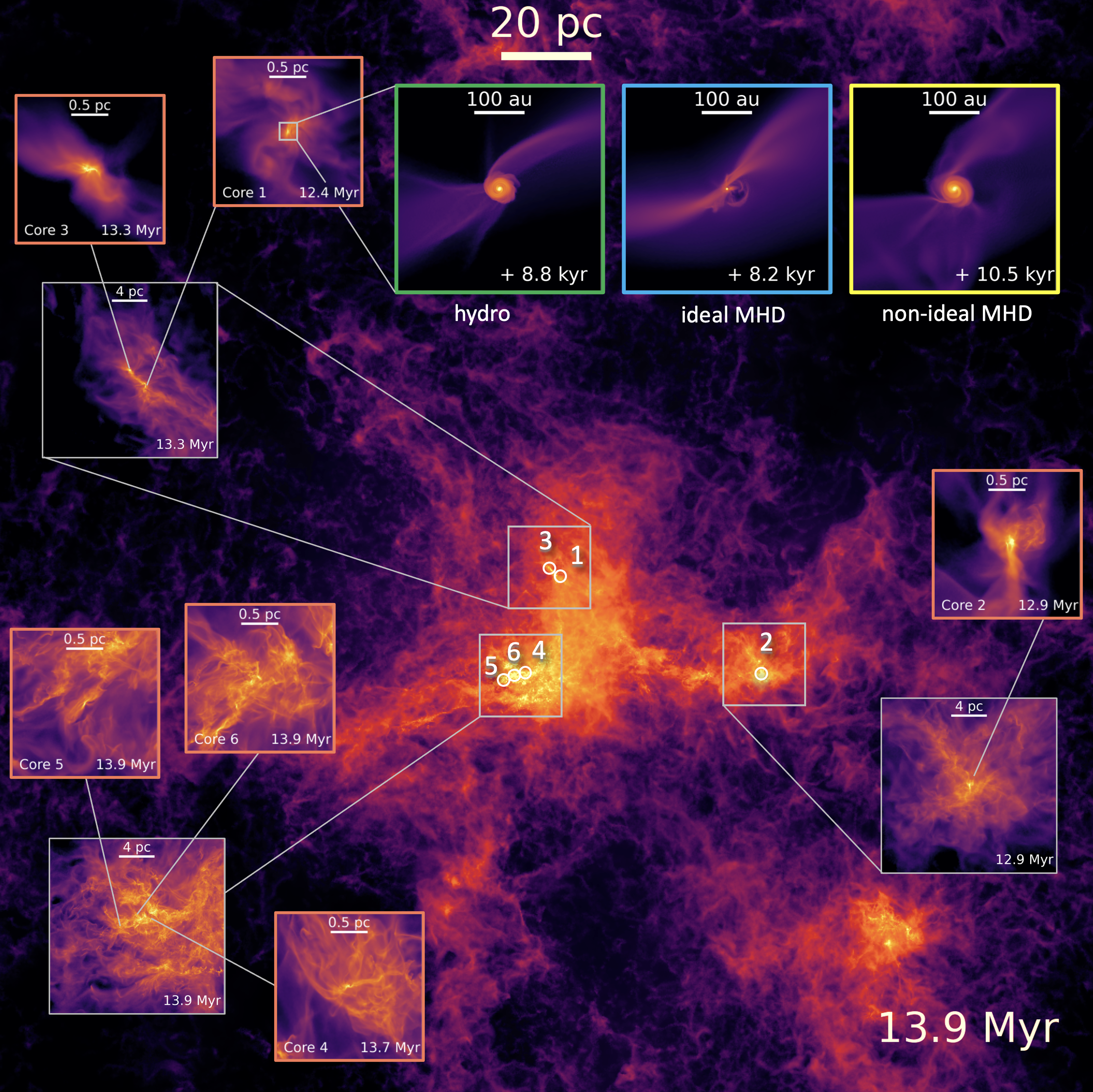}
    \caption{Schematic of the zoom-in simulation approach labeled by the respective size bars. The background image (20 pc) shows the turbulent gas distribution in the 256 pc supernova-driven box at the time (13.9 Myr) when the first six dense regions we investigate formed sinks (circles, see Sec. \ref{tracing}). The panels with scale-bars of 4 pc show the larger scale gas distribution and structure around the six individual cores (panels with 0.5 pc scale-bar) right before the respective sink formation. This is the starting point for the zoom-in simulations with a peak spatial resolution of $10^{-4}$ au. All six cores are simulated with hydrodynamics, ideal MHD, and non-ideal MHD, respectively. The three protostellar disk regions at the end of the simulations for Core 1 are shown in the 100 au panels. All panels show gas column density (different colorbar ranges) ranging from $\sim 10^{-4}$  g cm$^{-2}$ (darkest in background panel) to $\sim 10^{4}$ g cm$^{-2}$ (brightest in `hydro', `ideal' and `non-ideal' panels).}
    \label{fig:bigplot}
\end{figure*}

\label{sec:methods}
In this section, we explain the underlying methodology of our simulations concerning the simulation code as well as the modeling of physical processes such as cooling and non-ideal MHD.

\subsection{The (non-ideal) MHD simulation code {\small AREPO}}
The simulations in this work were performed using the moving-mesh magnetohydrodynamics code {\small AREPO} \citep{springel2010arepo,pakmor2016gradients,weinberger2020arepo}, which utilizes a finite-volume method on an unstructured moving Voronoi grid. Resolution elements (cells) automatically follow the fluid flow, which keeps the mass in every cell approximately constant, thereby sharing some of the advantages of smoothed particle hydrodynamics, like the automatically adaptive resolution and the reduction of advection errors in a manifestly Galilean-invariant formulation of hydrodynamics, while offering higher accuracy in the hydrodynamics. A refinement criterion that keeps the mass of cells within a factor of two around a (problem-dependent) predefined target mass is often used to resolve the relevant scales in a problem. This naturally yields a higher number of cells per volume in denser regions. However, the code is not bound to a fixed mass resolution in the entire simulation box, and cells can be split in a mathematically well-defined way, avoiding issues of particle splitting in finite-mass methods. We make use of this property in the refinement procedure (see Section~\ref{sec:simulations} for details) to \textit{further} increase the resolution in high-density regions. 

Magnetic fields were originally incorporated into {\small AREPO} in \cite{MHDArepo}, where the fluxes are calculated with the HLLD approximate Riemann solver \citep{miyoshi_kusano2005hlld}, with a fallback to more diffusive solvers if the HLLD returns a negative pressure. The divergence constraint for the magnetic field, 
$\nabla \cdot \bm{B} = 0$,
is guaranteed to stay fulfilled at all future times if true for the initial conditions at the analytical level (which follows from the form of the induction equation \eqref{eq:induction} below), but it is challenging to retain this property when the ideal MHD equations are discretized. Violations due to the build-up of discretization errors can cause instabilities if not controlled explicitly \citep{toth2000divergence}. 

Due to the moving-mesh nature of {\small AREPO}, divergence errors are most often simply the result of a change in discretization when interfaces are created or disappear in the Voronoi grid, and even simply due to the movement of the mesh-generating points. To address the divergence constraint, the Dedner cleaning scheme \citep{dedner2002cleaning} was first implemented into the code, but its need for a spatially constant cleaning speed, generally set to the current highest wave-speed in the simulation domain, makes it impractical for multi-scale simulations as the time-stepping of the entire box is then instantaneously affected by what is happening at one location in the simulation. Therefore, the standard method to treat the solenoidal constraint in {\small AREPO} has become the Powell divergence advection scheme \citep{powell1999cleaning}, as realized in the code by \citet{AREPOPowell}. We also utilize this method in this work.

Our implementation of non-ideal MHD is detailed in \cite{zier2024diffusion,zier2024hall}, although we only include ambipolar diffusion in this study as it is the dominant effect early in the collapse. The induction equation, modified to account for ambipolar diffusion, reads
\begin{equation}
\label{eq:induction}
    \frac{\partial \bm{B}}{\partial t} = \nabla \times (\bm{{\rm v} \times \bm{B}}) + \nabla \times \{ \eta_{\rm AD} [(\nabla \times \bm{B}) \times \bm{b}] \times \bm{b} \} \, .
\end{equation}
Here, $\bm{b} = \frac{\bm{B}}{\vert \bm{B} \vert}$ gives the direction of the magnetic field $\bm{B}$, and $\eta_{\rm AD}$ sets the strength of ambipolar diffusion, which in general depends on the local chemical composition, but in our case it is a function of magnetic field strength and density as we do not use non-equilibrium chemistry (see subsection \ref{subsec:coefficients}).

\subsection{Cooling and equation of state}
\label{subsection:temperature_treatment}
Previous studies that carried out zoom-in simulations of protostellar disks from larger scales have sometimes assumed optically thin cooling throughout the entire simulation and inserted relatively large sink particles \citep{kueffmeier2017zoom}. However, such a treatment is not appropriate on the smallest scales in our simulation, as we also model the phase of the first and second Larson core.

Instead of using a full radiation treatment for the temperature evolution or non-equilibrium chemistry, we also use equilibrium cooling to model the evolution of the supernova remnants and subsequent formation of prestellar cores. For the zoom-in regions within those cores, we  apply a barotropic equation of state, based on a density threshold. For the former, we adopt the same prescription as \cite{guo24snr}, except that we use a different (however, as in their case, fixed) mean molecular weight $\mu$ = 1.4 \citep[as used by ][ for their simulation with an equilibrium cooling formula]{micic2013cloud}, more appropriate for the neutral atomic ISM. As pointed out by \cite{micic2013cloud}, there are caveats related to this prescription even as a simplified equilibrium cooling formula, and it is likely not accurate even at significantly lower densities than the highest one we model using it ($n \sim 10^6$ cm$^{-3}$). A more accurate approach would of course need to follow the non-equilibrium abundances of different chemical species, which is, however, beyond the scope of this work \citep[see, e.g.,][for examples of a more complete description]{walch2015silcc,gatto2015ism}. \cite{guo24snr} use the low-temperature equilibrium cooling formula from \cite{koyama_inutsuka2002cooling} for $T < 10^{4.2}$ K:
\begin{equation}
\Lambda_{\rm lowT} (T) = \Gamma \cdot \left[10^7 {\rm exp}^{-\frac{1.184\cdot 10^5}{T+1000}} + 1.4\cdot 10^{-2}\, \sqrt{T} \,{\rm exp}^{-\frac{92}{T}}\right] \, {\rm cm}^3\,,
\end{equation}
with $\Gamma = 2\cdot 10^{-26} \, {\rm erg} \, {\rm s}^{-1}$ \citep[note our differing value, which set is as in ][]{micic2013cloud}. In the range 10$^{4.2} < T < 10^{8.15}$ K, we log-log interpolate the values in Table~2 of \cite{schure2009cooling}. Above 10$^{8.15}$ K, a fit from \cite{schneider_robertson2018cooling} is used:
\begin{equation}
    \Lambda_{\rm highT} (T) = 10^{0.45 \mathrm{log}(T) - 26.065} \mathrm{erg \, s^{-1} cm^3} .
\end{equation}
All of these cooling formulas and tables assume solar metallicity. 

With the additional heating term being constant and equal to $\Gamma$, the evolution of the total (thermal + kinetic) energy density of gas has the typical form (without the terms relating to magnetic fields)
\begin{equation}
\frac{\partial E}{\partial t} + \nabla \cdot [(E + P) \bm{\mathrm{v}}] = n \Gamma - n^2 \Lambda (T) \, ,
\end{equation}
with energy density $E$, thermal pressure $P$, gas velocity $\bm{\mathrm{v}}$ ($P$ is obtained via an ideal gas equation of state with $\gamma = \frac{5}{3}$), and number density $n$. We calculate the cooling rate of a cell with a standard implicit cooling algorithm. We set a temperature floor $T_{\rm floor} \approx  8$ K, which is generally reached in regions denser than $10^{-18}$ g cm$^{-3}$, so these are essentially treated as isothermal. 

For our zoom-in simulations, once a cell has reached the temperature floor and the cell has a density greater than $\rho_{\rm core} = 5 \times 10^{-18}$ g cm$^{-3}$, it is flagged as having entered the barotropic regime. The pressure of these cells is determined by their density alone, with the following equation of state \citep[which is a fit to the radiative calculations of ][]{wurster2018collapse}:
\begin{equation}
P = \rho c_{{\rm s},0}^2 \sqrt{1 + \frac{n} {n_1}}\left( 1 + \frac{n}{n_2}\right)^{-0.4}\left( 1 + \frac{n}{n_3}\right)^{0.37} ,
\label{eq: eos}
\end{equation}
where $n = \frac{\rho} {\mu \, m_{\rm p}}$ with $\mu \approx 2.381$, $m_{\rm p}$ being the proton mass,  and $c_{s,0}=0.22\,{\rm km\, s}^{-1}$ as well as  $n_1 = 2.0 \times 10^{10}\, {\rm cm}^{-3}$, $n_2 = 2.5\times 10^{14}\,{\rm cm}^{-3}$ and $n_3 = 1.0 \times 10^{20}\,{\rm cm\textbf{}}^{-3}$,  to match the default parameters in the utilized chemical library. This is the equation of state utilized in the simulations of isolated cloud-core collapse of  \citet{mayer25isolated}. The temperature floor above is chosen such that there is no pressure jump at the transition between the cooling- and barotropic regimes\footnote{Due to the jump in molecular weight there is however formally a jump in temperature, but since the barotropic regime denotes where we do not model the temperature independent of density anymore, this is of no concern.}.
Equation \eqref{eq: eos} includes a transition from the isothermal regime to adiabatic heating of gas (modeling the formation of the first Larson core) to the stage of hydrogen dissociation and proceeding to a second adiabatic phase that leads to the formation of the second Larson core at stellar-core densities. We note here that this equation of state represents a fit to the temperature evolution of the \textit{densest} fluid element in the simulation of \cite{wurster2018collapse}, and as such it systematically underestimates the temperature of any material that collapses later, which includes the entire protostellar disk.
With the above prescription for the temperature evolution, we obtain a basic model for the temperature evolution in the total range of densities in our simulation, which extends from the hot-phase ISM at $n < 10^{-3} {\rm cm}^{-3}$ to stellar core densities of $n > 10^{20} {\rm cm}^{-3}$.

\subsection{Non-ideal MHD coefficients}
\label{subsec:coefficients}
We utilize the chemical library {\small NICIL} to obtain the local ambipolar diffusion coefficient \citep{wurster2016nicil,wurster2021nicil2.0}. We assume a global cosmic-ray ionization rate of $\zeta_i = 10^{-17} {\rm s}^{-1}$ and a grain size distribution with grains divided into 5 logarithmically-spaced size bins distributed as a truncated MRN \citep{mathis1977mrn} profile from $a_{\rm min} = 0.1 \mu {\rm m}$ to $a_{\rm max} = 0.25 \mu {\rm m}$, referred to as 17mrn in \cite{mayer25isolated}. We then linearly interpolate the resulting table in density and magnetic field. While we did extend the use of the table to include a tri-linear interpolation that additionally uses the temperature of a cell, we have instead decided to keep the table barotropic. This is because the inside of the protostellar cores follows exactly this temperature treatment anyway, and this is the sole region of interest after the zoom-in has started. Using a two-dimensional table allows us to have a finer table for the same table file size. 

To reduce the stringent timestep requirements associated with the non-ideal MHD coefficients (which scale quadratically with the cell size, becoming very demanding at high resolution), we restrict the ambipolar diffusion coefficient similar to \cite{vaytet2018protostellar} and do not allow the timestep from ambipolar diffusion in any cell to fall below a prescribed fraction of the ideal MHD timestep: 
\begin{equation}
    \eta_{\rm AD} < \frac{1}{f_{\rm ni}} c_s r ,
    \label{ad_limit}
\end{equation}
where $c_s$ is the fast magnetosonic speed of a cell (the relevant wave-speed for the Riemann solver and thereby timestep of the cell), and $r = \left( \frac{3 V}{4 \pi} \right)^{\frac{1}{3}}$ is the volume-equivalent cell-radius. $f_{\rm ni}$ is a numerical parameter that sets the minimum non-ideal timestep in units of the ideal MHD timestep. We set $f_{\rm ni} = 0.125$. We only apply this limit if the non-ideal timestep falls below a prescribed value ($\approx$ 70 days), so that it should only noticeably affect the coefficient after the formation of the first hydrostatic core. Still, due to our use of this limit and the lack of Ohmic dissipation in our simulations, our non-ideal MHD scheme has a smaller effect than would be the physical effect for the assumed chemical model. In addition, cosmic-ray attenuation is substantial at densities typical of protostellar disks \citep{umebayashi_nakano2009cosmicrays} and the inclusion of this effect increases the strength of ambipolar diffusion and thereby alters disk evolution substantially \citep{nishio2025cosmicrays}. Note, however, that modelling uncertainties in the grain size distribution and related grain surface chemistry are also large \citep[see e.g][]{zhao2018grains}.

\section{Description of simulations}
Our simulation consists of three phases: A driving phase where we create a turbulent ISM via energy injection and without gravity, a tracing phase where we identify gravitationally collapsing regions, and finally zoom-ins where we look at one of these regions and simulate the process of protostar formation with three different magnetic field models.

\label{sec:simulations}
\subsection{Setup and driving}
Before the driving via supernovae, the initial conditions are fully homogeneous within the simulation domain, which is a periodic box of size (256 pc)$^3$. The initial number density is 1 cm$^{-3}$, with a molecular weight of $\mu = 1.4$ and a temperature of 4500~K. The initial magnetic field strength is set to $B_{\rm initial} = 0.3$ $\mu$G, parallel to the $z$-direction. These values, as well as the supernova explosion rate (see below), are based on models run by \cite{gatto2015ism} \citep[motivated by the Kennicutt-Schmidt relation][]{kennicutt1998schmidt} and for the magnetic field of \cite{pardi2017magnetic}, and are meant to model the conditions of the solar neighbourhood.

To obtain a turbulent ISM as initial conditions for the zoom-ins, the homogeneous box is then driven by injecting thermal energy ($10^{51}$ erg) into the 128 nearest neighbours of random cell positions. During this driving, gravity is switched off. Since we do not simulate actual star particles, the rate of injections in the whole box is characterized by
\begin{equation}
    f_{\rm box} = f_{\rm SN} \times \left( \frac{{\rm M}_{\rm box}}{10^6 {\rm M}_\odot} \right) \rm{Myr}^{-1} \, ,
\end{equation}
where we set $f_{\rm SN} = 6.0$. We have $\sim 5\times 10^5$ M$_\odot$ of mass and drive for $\sim 10$ Myr, so $\sim 30$ supernovae typically happen over the driving time, and the homogeneous initial conditions are completely erased (see left panel of Figure \ref{fig:after_driving} in Appendix \ref{appendix_after_drving}).

The driving and evolution up to the zoom-in is always performed with ideal MHD, since changing the driving procedure would already lead to very different initial conditions, whereas we are interested in the differences between the star formation processes for the three magnetic field models (no magnetic fields, ideal MHD, ambipolar diffusion). For the driving, this mostly concerns differences between the cases of magnetic fields and no magnetic fields, as the effects of ambipolar diffusion likely only becomes important on smaller scales than are resolved in the driving phase ($\sim 0.05$ pc minimum cell size). Note that we do not include shear forces to model galactic rotation in our simulations \citep[see e.g.,][]{kim_ostriker2017tigress}, and as such, this further (fully solenoidal) driver of turbulence is absent in our current model \citep[however, there is an implementation of the shearing box approximation in {\small AREPO} compatible with MHD and self-gravity as detailed in ][]{zier2022shearing, zier2022shearingIdealMHD, zier2022shearingGravity}.

During the driving phase, we use a mass refinement with $M_{\rm max} = 0.0375\, {\rm M}_\odot$, which serves as the \textit{maximum} target mass of cells in the box until the zoom-in. This is easily sufficient to resolve the Sedov-Taylor phase of the supernovae \citep[see e.g.,][]{kim2015ism}, and thermal injection is therefore a valid injection mechanism.

\subsection{Identification of zoom-in targets (tracing)}\label{tracing}
\begin{table*}
\begin{tabular}{c|c|c|c|c|c|c|}
      \hline
      Simulation phase & $M_{\rm max}$ [M$_\odot$] & $M_{\rm min}$ [M$_\odot$] & $n_J$ & $n_{J}^{\rm iso}$ & $r_{\rm min}$ [au]& Additional criteria\\
      \hline
      Driving & $3.75\times 10^{-2}$ & $3.75\times 10^{-2}$ & - & - & - & $r_{\rm max}$ = 2 pc \\
      Tracing & $3.75\times 10^{-2}$ & $6.0\times 10^{-5}$ & 32 & 8 & 25 & Sink particles at $\rho > 5 \times 10^{-16}$ g cm$^{-3}$\\
      Zoom-in & - & $3.33\times 10^{-7}$ & 32 & 8 & - & De-refinement outside zoom-region\\
\hline
\end{tabular}
\caption{Refinement criteria used in the different phases of our simulations. $M_{\rm max}$  and $M_{\rm min}$ are the upper and lower limits for the mass of a cell (within a factor of 2). $n_J$ and $n_{J}^{\rm iso}$ are the number of cells per Jeans length per dimension with the usual Jeans criterion and the isothermal Jeans criterion, respectively (see text, the Jeans refinement is the reason for differing cell masses). $r_{\rm min}$ and $r_{\rm max}$ are the minimum and maximum allowed volume-equivalent cell-radius, respectively.}
\label{RefinementTable}
\end{table*}
Since we aim to obtain a mass resolution better than $10^{-6}\, {\rm M}_\odot$ while starting with more than $10^5 {\rm M}_\odot$ of mass in the simulation domain, a refinement that aims to keep the mass of all cells approximately equal and constant (which is well-adapted to the quasi-Lagrangian nature of {\small AREPO}) would lead to a very large number of cells ($> 10^{11}$) while making little difference for the evolution of the densest regions. Instead, we utilize Jeans-type criteria, similar to what has been used in simulations of molecular clouds using {\small AREPO} \citep[e.g.,][]{dhandha2024decaying}. 

After gravity is enabled and the supernova injections are turned off at the end of the driving phase, we add an `isothermal Jeans criterion', i.e., with a constant sound speed of $0.2\,{\rm km\, s}^{-1}$ (such that this modified Jeans length is only a function of density). We require that the isothermal Jeans length is always resolved by at least $n_{J}^{\rm iso} = 8$ cells per dimension. We supplement this with a standard Jeans criterion, where the local Jeans length at any point is required to be resolved with a minimum of $n_{J}=32$ cells per dimension, provided this is not already fulfilled by the isothermal refinement. After the end of driving, the mass refinement alone is sufficient to resolve the Jeans length almost everywhere. 

We limit the refinement in the whole box with a global volume-equivalent cell radius of approximately 25 au ($\approx$ one eighth of the isothermal Jeans-length at the sink-creation density, see below) and a minimum mass of $6.0\times 10^{-5} {\rm M}_{\odot}$ to speed up the calculation\footnote{This implies that the number of cells per Jeans length is lower once this minimum is reached, and this implies that relatively rapid refinement happens at the start of the zoom-in. However, the Truelove criterion \citep{truelove1998fragmentation} of 4 cells per Jeans length per dimension is never violated.}. Since the cell size decreases rapidly with increasing density for Jeans-refinement, and stark density contrasts occur in our simulations, we additionally require that the radii of neighbouring cells may only differ by a factor of two.

To follow the evolution of the full box for as long as possible, we utilize the standard implementation of sink particles in {\small AREPO}. We insert them at a density of $\approx 5 \times 10^{-16}$ g cm$^{-3}$, if other standard tests for formation are also passed (gravitational boundedness, converging flow in terms of velocity and acceleration, and the cell has to be a local potential minimum). We observe a large number of structures which reach densities above $10^{-18}\,{\rm g\, cm}^{-3}$ but turn out to be transient \citep[consistent with the results of][]{offner2025starforgecores}, while the density threshold used here appears to be robust in identifying actual collapsing regions. 

The sinks accrete material in their accretion radius (set to $\approx 200$ au) above the sink formation density, also taking momentum and magnetic fields from such cells, proportional to the accreted mass. Sink particles are allowed to merge if one enters the accretion radius of another. We should note that the details of the sink particles essentially do not matter for our current work, because we are only interested in where and when they form to show us where to zoom in while making sure that we pick regions which are sufficiently far apart to prevent any influence of neighboring sink particles. We do not include any feedback processes from formed sinks, which would in reality affect the later evolution of the rest of the box. This would happen for sure at the point of the first supernova, although protostellar feedback in the form of radiation and jets would have an effect even sooner \citep{neralwar2024feedback}. The first sink in our simulations forms approximately 12.4 million years after the end of the driving, and the last one we zoom in on forms after about 13.9 million years, but we have been able to run the evolution much further. 
\begin{figure}
    \centering
    \includegraphics[width=1\linewidth]{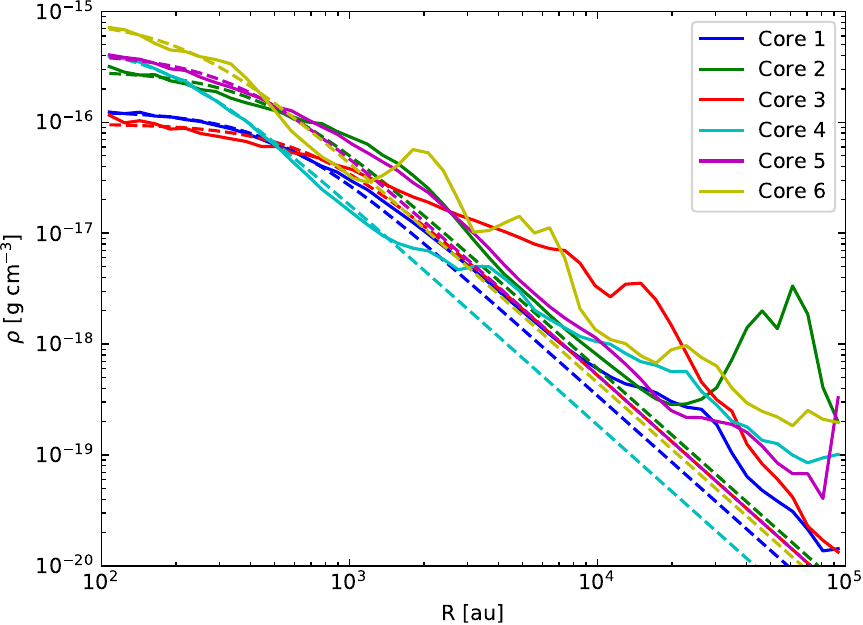}
    \caption{Density profiles of the six cores (solid lines) and fitted Bonnor-Ebert spheres (dashed lines) at the start of the zoom-in simulations (see also the panels with a scale-bar of 0.5 pc in Fig. \ref{fig:bigplot}). The centres of the cores follow a Bonnor-Ebert distribution. In the outer parts, the cores are not spherical (see also Fig. \ref{fig:initial_cores_cut}).}
    \label{fig:core_profile}
\end{figure}
\subsection{Zoom-in simulations}
After a region of interest has been identified as a possible zoom-in region, we go back to a snapshot before sink formation and add a tracer fluid to the relevant snapshot around the region where the sink has formed\footnote{We have checked that starting the zoom-in earlier does not significantly affect the results, see Appendix \ref{appendix_checks}.}. In this region, instead of forming a sink particle, we resolve the protostellar collapse up to the formation of the second Larson core, as cells in the zoom-in region are allowed to refine further (utilizing both Jeans criteria) up to a minimum mass of $3.33\times 10^{-7} {\rm M}_{\odot}$. This corresponds to the mass resolution used in the isolated runs in \cite{mayer25isolated}, and is easily sufficient to resolve the Jeans mass up to the formation of the second Larson core. 
\begin{figure*}
    \centering
    \includegraphics[width=1\linewidth]{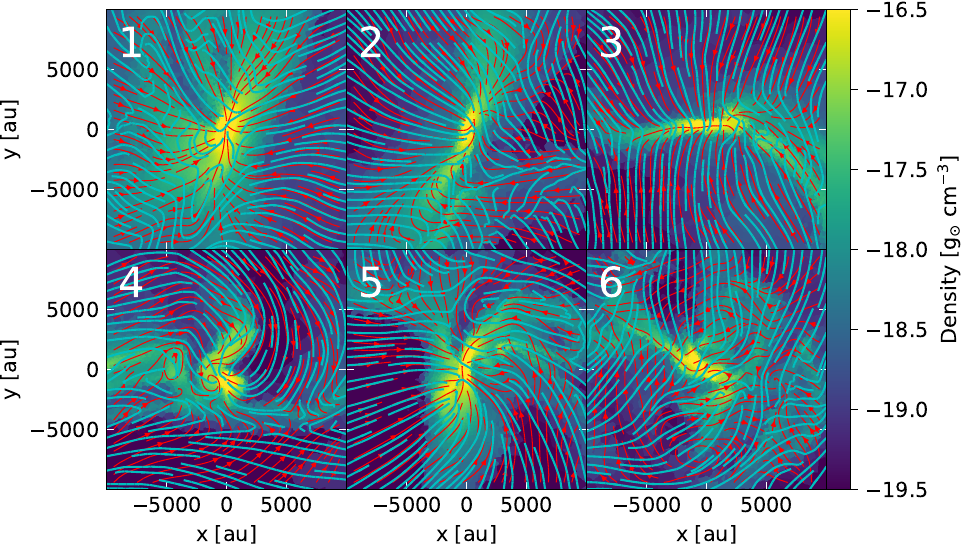}
    \caption{Slices (zero width) through the initial cores 1 to 6 from top left to bottom right (see 0.5 pc panels in Fig. \ref{fig:bigplot}). We show density (colour), in-plane magnetic field direction (cyan streamlines), and velocity direction (red lines and arrows), where the latter is shifted to the centre of mass velocity of the core. The cores display a highly complex kinematic and non-spherical morphological structure. In general, there is no clear correlation between density, magnetic field lines, and flow direction. Core 3 (top right) shows the most ordered configuration with perpendicular flows onto the elongated core filament and perpendicular field lines.}
    \label{fig:initial_cores_cut}
\end{figure*}
\begin{table*}
\begin{tabular}{c|c|c|c|c|c|c|c|c|c|}
      \hline
      Core id & $M$ [M$_{\odot}$] & $\rho_{\rm mean}$ [10$^{-17}$ g cm$^{-3}$] & $r_{\rm eff}$ [AU] & $L$ [M$_{\odot}$ AU km s$^{-1}$]& $j$ [AU km s$^{-1}$] & $\sigma_v$ [km s$^{-1}$] & $B_{\rm mean}^{\rm abs}$ [$\mu$ G] & $\sigma_B$ [$\mu$ G] \\
      \hline
1 & 0.95 & 1.19 & 2248.18 & 58.71 & 61.51 & 0.3 & 173.16 & 183.18 \\ 
2 & 1.22 & 1.41 & 2301.65 & 179.07 & 147.19 & 0.25 & 246.15 & 351.15 \\ 
3 & 3.36 & 1.11 & 3497.33 & 3179.01 & 945.41 & 0.36 & 203.94 & 203.8 \\ 
4 & 1.05 & 0.96 & 2489.58 & 574.86 & 548.46 & 0.42 & 117.15 & 330.9 \\ 
5 & 1.92 & 1.3 & 2759.75 & 726.26 & 378.27 & 0.33 & 252.58 & 348.74 \\ 
6 & 2.04 & 1.58 & 2634.92 & 449.95 & 220.9 & 0.48 & 188.79 & 460.38 \\ 
\hline
\end{tabular}
\caption{Physical properties of the initial cores, which are defined as all material with $\rho > \rho_{\rm core} = 5 \times 10^{-18}$ g cm$^{-3}$ and within a box of size 0.1 pc around the sink fomration position. $r_{\rm eff}$ is the volume-equivalent radius of the core, $L$ the angular momentum, $j$ the specific angular momentum, $\sigma_v$ the velocity dispersion, $B$ the mean magnetic field, and $\sigma_B$ the magnetic field dispersion. The velocity dispersion is calculated as square root of the mass-weighted average of the squared velocity in the centre-of-mass frame of the core, while the mean magnetic field strength is volume-weighted and the dispersion calculated as the square of the volume-weighted average of $\vert \bm{B} - \bm{B}_{\rm mean} \vert^2$, where $\bm{B}$ is the magnetic field of a given cell.}
\label{Initialcores_table}
\end{table*}

To save computational time, the resolution outside of this region is slowly decreased, as they are now only of interest as a boundary condition. Depending on the simulation, we either remove the magnetic field or switch on ambipolar diffusion at this point. We will occasionally use the terms ``hydro'' for the simulations without magnetic fields, ``ideal" for those with ideal MHD, and ``non-ideal" for the ones with ambipolar diffusion, and the shorthands ``h/i/n[\textit{zoom-target number}]". For example, ``i4'' refers to region 4 simulated with ideal MHD. The simulations without magnetic fields should be seen mostly as a point of reference for the evolution of the same core without any magnetic effects. Note that in this case, we remove magnetic support from the hydro cores, and the initial conditions are not strictly self-consistent.

Generally, we run the simulations until the formation of the second core, although this was not possible for simulation n6 because material was collapsing at the edge of the zoom-in region before the main zoom-target, which is, unfortunately, a complication unavoidable in general in our approach. However, we were still able to run this region until the second collapse had almost begun ($\rho_{\rm central} \approx 2\times10^{-9}\,{\rm g\, cm}^{-3}$). The cores we chose are the first isolated protostellar cores which form sufficiently isolated from each other in the box (in total, there are more than 50 sinks, and more have formed but have been removed by mergers); we excluded one of them simply because it showed relatively similar behaviour to region 4.

Figure \ref{fig:bigplot} shows a general overview of our simulations and displays the structure of the whole box as well as of individual zoom-in regions. It also includes one example of protostars resulting from the zoom-in procedure (coloured boxes in the top centre).

\section{Properties of selected cores / regions}
\label{sec:cores}
\subsection{Overall properties}
\begin{figure*}
    \centering
    \includegraphics[width=1.0\linewidth]{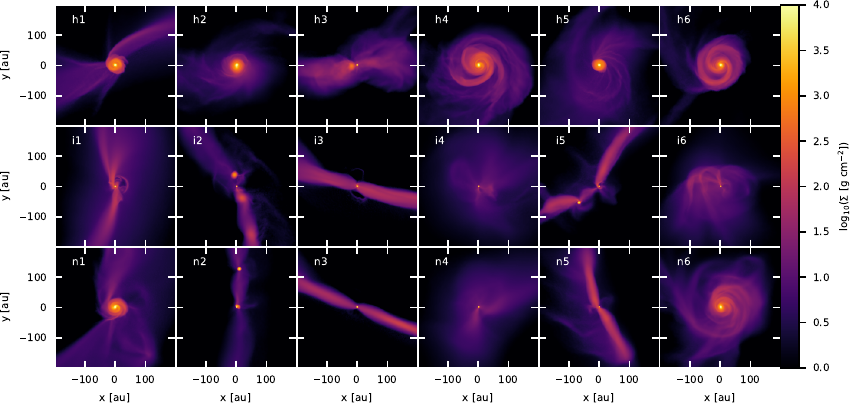}
    \includegraphics[width=1.0\linewidth]{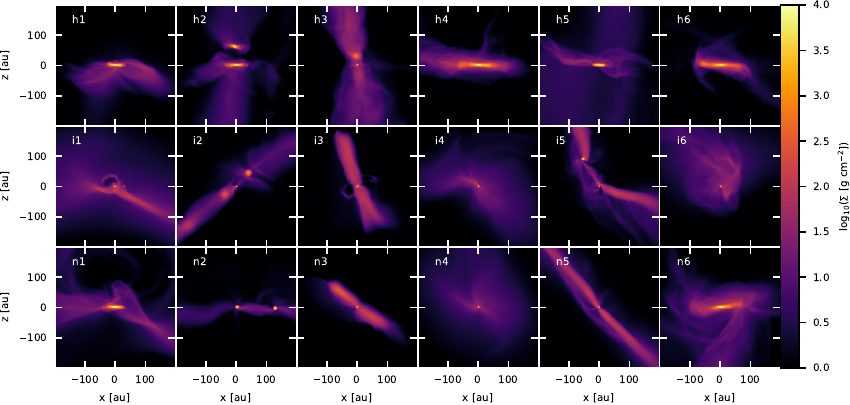}
    \caption{Density projections at the end of each zoom simulation of the collapse of cores 1 to 6 (left to right), in face-on (three top rows) and edge-on directions (three bottom rows), as defined by the angular momentum of the first core. For each projection, we show the hydrodynamic (h, top), ideal MHD (i, middle), and non-ideal MHD (n, bottom) simulations.  On this scale, extended disks up to 100 au form in all hydro simulations except for h3, while there are disks for ideal MHD. With ambipolar diffusion, extended disks form in n1 and n6.}
    \label{fig:final_cores}
\end{figure*}
As a first look at the prestellar cores, we display the spherically averaged density profiles in Figure \ref{fig:core_profile}, along with a fit\footnote{We perform a linear and not a logarithmic regression, so it is biased to fit closer in the inner, denser region.} in the form of a Bonnor-Ebert sphere, which was first inferred as a typical shape of the density profile of molecular cloud cores by \citet{alves2001bonnorebert,tafalla2002cores}:
\begin{equation}
    \rho (r) = \frac{\rho_0}{1 + \left( r / r_0\right)^2} \, ,
\end{equation}
with the free parameters being the central density $\rho_0$ and the characteristic radius $r_0$. This profile describes a singular isothermal sphere for $r \gg r_0$, but with a finite density at $r=0$, and has often been used as an initial condition for star formation \citep[e.g.,][]{xu2021formationI,zier2021BonnorEbert}. 

Looking at Figure \ref{fig:core_profile}, we can see that the inner parts ($< 1000$ au) are in fact nicely fit with such a profile, but the density is usually much higher than in the Bonnor-Ebert sphere at larger radii, and appears to start merging with the ISM at a distance of $\sim 0.1-1$ pc from the centre. This is consistent with the typical size of dense cores \citep{myers_benson1989cores,andre2014cores}. Values of $r_0$ range from $\approx 185$ au (core 4) to $\approx 735$ au (core 3). These values correspond to the pre-stellar core models studied in \cite{keto_caselli2010cores}, who pointed out that the size of the region with a flat density profile (our $r_0$) is given by the product of the sound speed and the free-fall time at the central density (see their equation 2). Our range of values are also in line with the physical structure of the prototypical pre-stellar core L1544, where $r_0 \sim 500$ au \citep{keto_caselli2010cores,caselli2019l1544}. \cite{caselli2022l1544} also noted that the 3D structure of the simulation used for comparison with data had density profiles which could be reduced to BE spheres (see Figure D.1).

Column density projections of the cores at the beginning of the zoom-in are shown as insets with orange frames along with core numbers in Figure \ref{fig:bigplot}, while slices with velocity vectors and magnetic field lines are shown in Figure \ref{fig:initial_cores_cut}. For the latter, we first calculate the inertia tensor of each core; the slices are then shown in the plane perpendicular to the axis with the greatest eigenvalue (i.e., moment of inertia), as this should be the plane containing the most material. We define the core as all material in the selected zoom-region with a density higher than $\rho_{\rm core} = 5 \times 10^{-18}\,{\rm g\, cm}^{-3}$, a value we adopt from \cite{yang2025zoom}. Even at first glance, we observe that they are neither spherical nor isolated. In Table \ref{Initialcores_table}, we list some relevant quantities of the initial cores. We note that these quantities are significantly less useful in characterizing the initial conditions for star formation than in isolated collapse scenarios, as it is not expected that all of the material will fall onto one central object \citep[as shown, e.g., by][where the dependence of fragmentation on the properties of the core is discussed]{walch2012imf}, and we shall in fact see that the core collapses into multiple hydrostatic cores in some of our simulations. Some material may also become unbound from the core.

\begin{figure*}
    \centering
    \includegraphics[width=1.0\linewidth]{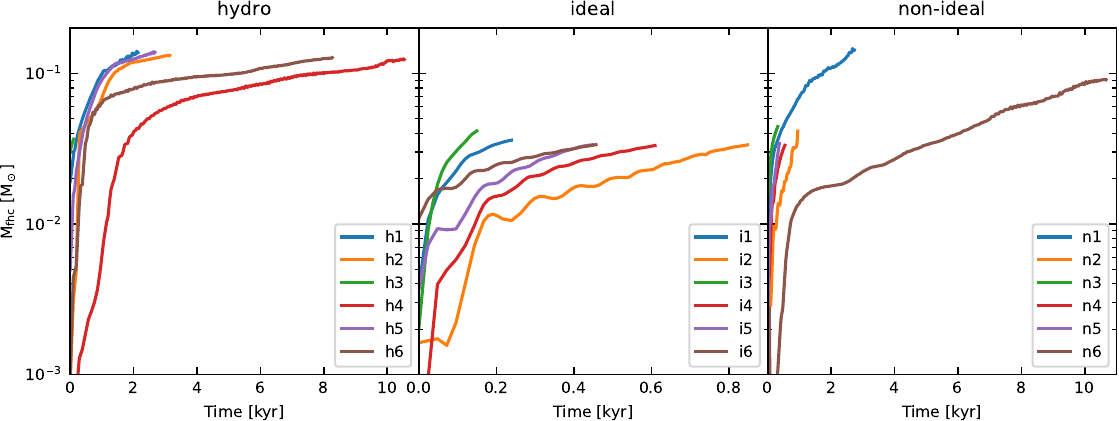}
    \caption{Masses in the hydrostatic cores for all 6 zoom-targets as a function of time from first to second core formation for the hydrodynamical (h, left), ideal MHD (i, middle), and non-ideal MHD (n, right) zoom simulations. There is a large variation in the time it takes for the forming protostar to transition from the first to the second core, even for the same magnetic field model. In general, the time is far shorter for ideal MHD (not the different time axis), and the masses are slightly lower. The mass for n6 decreases at the beginning  as it briefly drops below our density threshold for the definition of the first core. The cores forming discs are the most massive ($M_{\mathrm{fhc}} \sim 0.1 {\rm M_\odot}$).}
    \label{fig:mass_in_core}
\end{figure*}

\subsection{Morphology}
Our cores strongly depart from sphericity, and are also all connected to some degree to larger structures of their parent molecular cloud. Notably, cores 2, 3, 4, and 5 have the appearance of a dense substructure of a filament as they are elongated along one axis when viewed on a scale of 1-2 pc (six smallest insets in Figure~\ref{fig:bigplot}). All cores are also highly irregular on sub-pc scales (Figure~\ref{fig:initial_cores_cut}), where we can also see that the magnetic field is mostly perpendicular to the elongated dense filamentary structures (radius < 0.01 pc) and the inflow is proceeding along the field lines. However, in the densest regions, the magnetic field starts to align more with the filament instead. This trend is a general phenomenon seen both in simulations and observations \citep{gomez2018magnetic,anez2020magnetic}. 

In cores 2 and 4 there are large velocity jumps across the dense region which indicates that these in particular are likely formed in a shock front (originating ultimately from a supernova blast during the driving). As such, we expect these cores in particular to behave very differently from the homogeneous cores, which have often been studied in simulations of cloud-core collapse.

\section{Analysis of zoom-in regions}
\label{sec:results}
\subsection{General morphology and occurrence of disks}

We start by describing the overall picture of the hydrostatic cores and disks formed in our zoom-in simulations. Column density projections of the inner 200 au are shown in Figure~\ref{fig:final_cores} for all runs, both face-on (upper panels) and edge-on (lower panels). ``Face-on'' is here defined via the angular momentum of the first hydrostatic core, taken as all material with a density $\rho > \rho_{\rm fhc} = 10^{-11}$ g cm$^{-3}$. In all cases, the last snapshot is shown, which is within 10 years of the formation of the second Larson core, except for n6. It is immediately clear that all hydro simulations have formed disks of a size of more than 10 au, except for h3 (in this case, the first collapsing fragment simply had very low angular momentum). The simulations with ideal MHD are completely different in that no disks are visible at all, and instead, only approximately spherical hydrostatic cores form. 

For non-ideal MHD, there are 3 disks: Very prominently in n1 and n6 (in this case approaching 100 au from visual inspection), and a small disk in n2, which seems to form as the result of the merger of two hydrostatic cores.
The cores or disks show a large variety of morphologies, in particular in how they connect to their surroundings. For example, while the filament lies approximately in the plane of the disk in the hydro simulation of core 1, it is instead perpendicular to the disk for core 2. Core 2 is also notable because the filament has fragmented into two hydrostatic cores in all models, while in core~5 this is only the case with ideal MHD. Core~3 appears to be located at a strong shock front, which makes it hard to form a disk before the second collapse, even in the hydro case (as the hydrostatic core is quickly compressed).
\begin{figure*}
    \centering
    \includegraphics[width=1.0\linewidth]{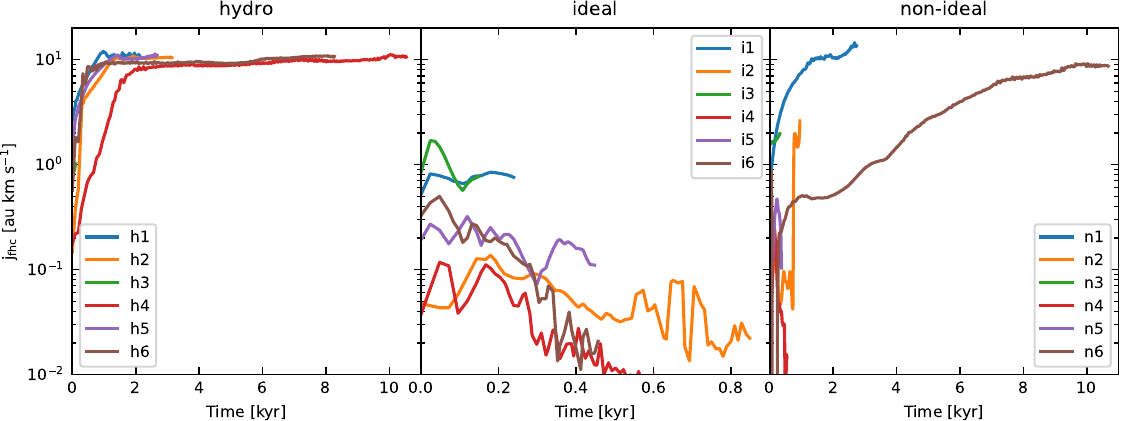}
    \caption{Specific angular momentum in the first cores as a function of time since their formation (similar to Fig. \ref{fig:mass_in_core}). The value reaches a plateau of $\sim 10$ au km s$^{-1}$ in the hydro case (left) and the two discs in the non-ideal MHD (n1 and n6), right panel. This might originate from the transport of angular momentum to low-density regions of the disk. The angular momentum declines for the ideal MHD zoom simulations, consistent with the effect of outflows.}
    \label{fig:j_in_core}
\end{figure*}

\begin{figure*}
    \centering
    \includegraphics[width=1.0\linewidth]{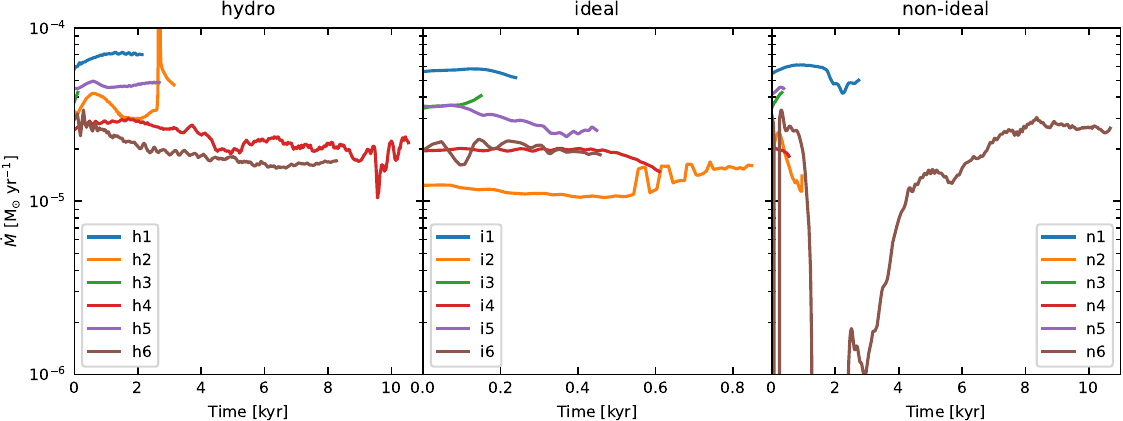}
    \caption{Instantaneous net flow-rates through the surface of a sphere with radius 200 au centred on cores 1 to 6. The $x$-axis measures time from the formation of the first hydrostatic core (same ordering as in Figs. \ref{fig:mass_in_core} and \ref{fig:j_in_core}). All cores have comparable mass-accretion rates of the order $10^{-4}$ to $10^{-5}$ M$_\odot$ yr$^{-1}$ on this scale. The sudden uptick in the model h2 (left panel) occurs as a secondary core enters the 200 au radius.}
    \label{fig:inflow}
\end{figure*}

\subsection{Masses, angular momentum and accretion rates}

In Figure \ref{fig:mass_in_core}, we show the mass of the first core over time (ending close to the collapse into a second Larson core), starting from the time of its formation in each simulation. The hydro simulations generally reach masses of larger than 0.1 M$_{\odot}$ before the mass is sufficient to proceed to second collapse. On the other hand, the ideal MHD simulations and the non-ideal ones that do not form a disk initiate the formation of the stellar core at a lower mass of the first hydrostatic core. The ideal MHD cores always collapse quickly (< 1 kyr), while in the hydro case, this is generally prevented for a much longer time (and can, as such, also grow to larger masses), on the order of ten times longer. This is generally connected to a larger specific angular momentum in the first hydrostatic core, which is shown as a function of time from first core formation in Figure \ref{fig:j_in_core}. In the case of ideal MHD, this quantity actually generally decreases over time, which is consistent with the removal of angular momentum by magneto-rotational outflows while accretion leads to a gradual gain in mass in the hydrostatic core (see subsection \ref{subsection:outflows} below). In the hydro cases, it reaches a maximum value, due to transport of angular momentum outwards in a manifestation of disk self-regulation and likely as a result of gravitational torques \citep[see, e.g., the discussion in][]{xu2021formationI}, so that angular momentum is contained in gas of lower density than our threshold for the hydrostatic core. 

We note that this is consistent with the structure of the disks as seen in Figure~\ref{fig:final_cores}, where the hydro disks generally show extended spiral arms. The non-ideal MHD simulations, which form large disks (n1 and n6), also show similar maximum values of the maximum specific angular momentum (note again that n6 has not yet reached second collapse, but we expect it to reach a similar value in the plateau). These two simulations also form first hydrostatic cores already at a relatively high specific angular momentum compared to the others. The exception is, again, core~3, where the collapse proceeds quickly in all cases despite high initial first core angular momentum. It appears that the collapse is much delayed by magnetic forces in this case. 

Figure~\ref{fig:inflow} (inflow) shows the instantaneous net mass flow rate (where the outflow rate is subtracted) over time, again since the formation time of the first core, on a shell of radius 200 au. Despite large morphological differences between the six cores and vastly different resulting structures on au scales, the overall inflow rate at this radius is generally similar (of the order $10^{-4}$ to $10^{-5}$ M$_\odot$ yr$^{-1}$ on this scale). However, we do observe a reduction of the inflow rates with the inclusion of magnetic fields with respect to hydro, consistent with the slowing of infalling material by magnetic tension and pressure forces. The inflow rate at this scale is consistent within a factor of a few with numerical estimates of accretion onto a protostar and disk/core system \citep{xu2021formationI}. The accretion is far from spatially homogeneous (both between models and within one model), which we further discuss in subsection \ref{subsection:outflows} below. 

\begin{figure*}
    \centering
    \includegraphics[width=0.9\linewidth]{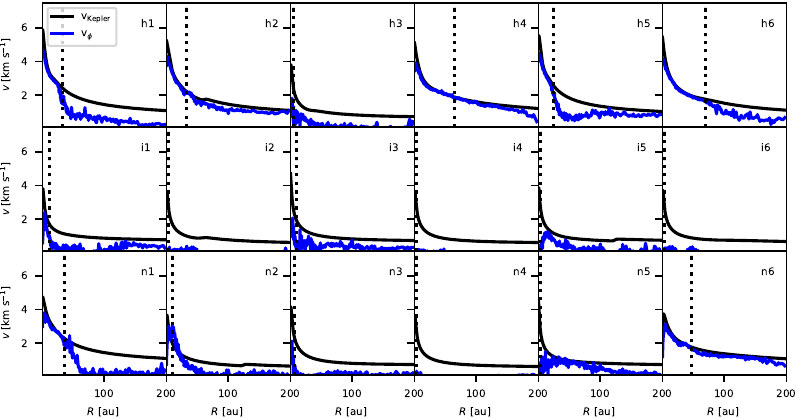}
    \caption{Final Keplerian (black) and tangential velocity (blue) as a function of radius for the six simulations (left to right) for the hydrodynamical (h, top), ideal MHD (i, middle), and non-ideal MHD (n, bottom) simulations. The Keplerian velocity is calculated assuming all matter inside the given radius is spherically distributed. Dotted vertical lines indicate the disk size estimated with the orbit/geometric method described in Sec. \ref{subsec:disk_rotation}. In many cases, the disk size estimates capture the deviation from Keplerian velocity (e.g., h1, h5, h6, n1). In other cases, the discs follow close Keplerian rotation to larger radii (e.g., h2, h4, n6). If the departure from Keplerian was used as a disc size criterion, h4 and n6 would be the largest discs in the sample. This would also agree with the visual impression from Fig. \ref{fig:final_cores}.}
    \label{fig:velocity_profiles}
\end{figure*}

\subsection{Disk sizes and radial profiles of disk quantities}
\label{subsec:disk_rotation}

\begin{table*}
\begin{tabular}{c|c|c|c|c|c|c|c|c|c|c|c|c|c|c|c|c|c|c|}
\hline
      & h1 & h2 & h3 & h4 & h5 & h6 & i1 & i2 & i3 & i4 & i5 & i6 & n1 & n2 & n3 & n4 & n5 & n6\\
\hline
$M_{\rm disk}$ [M$_\odot$] & 0.21 & 0.18 & 0.04 & 0.25 & 0.18 & 0.23 & 0.04 & 0.03 & 0.04 & 0.03 & 0.03 & 0.03 & 0.21 & 0.04 & 0.04 & 0.03 & 0.03 & 0.15 \\  
$M_{\rm disk}$ / $M_{\rm fhc}$ & 1.54 & 1.37 & 1.0 & 2.06 & 1.33 & 1.84 & 1.07 & 1.0 & 1.0 & 1.0 & 1.02 & 1.0 & 1.44 & 1.0 & 1.0 & 1.0 & 1.0 & 1.69 \\ 
$R_{\rm eff}$ [au] & 32 & 37 & 4 & 68 & 24 & 72 & 11 & 2 & 7 & 2 & 9 & 2 & 36 & 4 & 3 & 2 & 3 & 51 \\  
\hline
\end{tabular}
\caption{Disk masses $M_{\rm disk}$, ratios of disk mass to the mass $M_{\rm fhc}$ of the first hydrostatic core and disk radii, with the definitions detailed in subsection \ref{subsec:disk_rotation}.}
\label{DisksizeTable}
\end{table*}

To define the radius and material of protostellar disks, previous studies have often chosen either a density-based definition, where all material above a certain density threshold is counted as part of the disk \citep[e.g.,][]{yang2025zoom} or they have defined it kinematically, by computing an azimuthally averaged rotation profile and identifying either a sharp drop in the angular velocity \citep{xu2021formationI} or the radius where the tangential velocity falls below some fraction of the Keplerian velocity \citep{kueffmeier2017zoom}. We use a criterion mostly based on \cite{bate2018diversity}, which identifies individual resolution elements (instead of radial bins) as part of the disk, but is not solely based on density. A particle is counted as part of the disk if it has an instantaneous orbital eccentricity of $e < 0.3$ with respect to a reference frame consisting of the sink particle in the centre of the disk and all resolution elements that have been previously assigned. In their case, the procedure is started at the sink and then proceeds outwards through all cells sorted by increasing radius. 

Since we do not have a sink particle to start from, we instead make all cells in the first hydrostatic core (defined as before through the density cut $\rho_{\rm fhc} > 10^{-11}$ g cm$^{-3}$)  part of the disk initially. We only allow for material with $\rho_{\rm disk} > 8 \times 10^{-14}$ g cm$^{-3}$ to be part of the disk, which excludes most outliers that are really part of the envelope. The size of the disk is then defined as the corresponding radius of the area $A_{\rm plane}$ in the plane of rotation (defined by all identified disk material) that is covered by the disk, considering material within a scale height of 5~au. Specifically, we define the corresponding effective radius as $R_{\rm eff} \equiv \sqrt{\frac{A_{\rm plane}}{\pi}}$. 

The disk masses, the ratio of disk mass to first hydrostatic core mass (recall that the latter is always defined as disk), and disk sizes resulting from these definitions are listed in Table~\ref{DisksizeTable}. We can see that the results match quite well with what would be expected by visual inspection of Figure~\ref{fig:final_cores}. Except for region 1, there is essentially no disk material in the ideal MHD runs ($M_{\rm disk}$ / $M_{\rm fhc}$ < 1.1), which is also reflected in small radii. The disks formed in the hydro runs are instead larger than 10 au in all cases except for h3, while in the non-ideal MHD case, the disks of n1 and n6 are significantly larger than the others. 
\begin{figure*}
    \centering
    \includegraphics[width=0.9\linewidth]{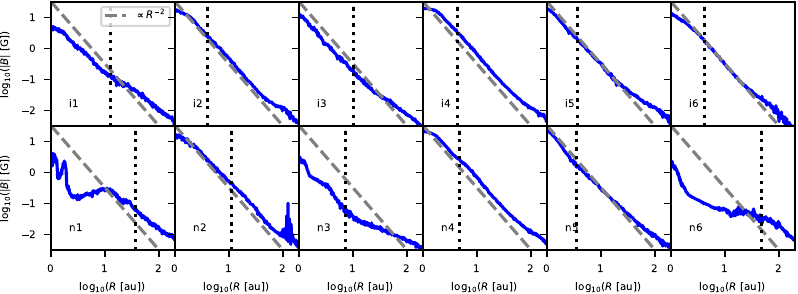}
    \caption{Absolute value of the magnetic field as a function of radius on a logarithmic scale for the final cores 1 to 6 from left to right for the ideal MHD (i,top) and non-ideal MHD (n, bottom) zoom simulations. The dotted vertical lines show the disc sizes as in Fig. \ref{fig:velocity_profiles}. The dashed lines indicate a power-law scaling with a slope of -1.5. The ideal MHD and non-ideal MHD cores, which do not form discs, approximately follow this scaling with peak central values of $\sim 10$ G. Discs (n1,n6) strongly deviate from this scaling in the disc region.}
    \label{fig:babs_profiles}
\end{figure*}
\begin{figure*}
    \centering
    \includegraphics[width=0.9\linewidth]{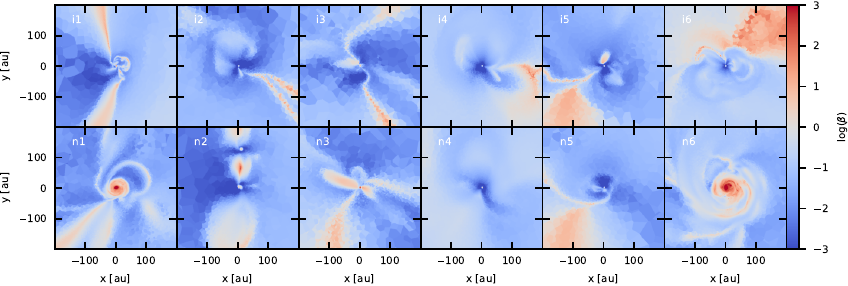}
    \includegraphics[width=0.9\linewidth]{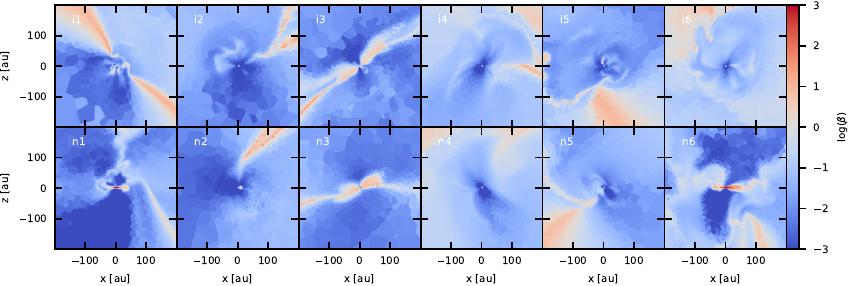}
    \caption{Plasma beta ($\beta = P_{\rm therm} / P_{\rm mag}$) at the end of each ideal (i) and non-ideal (n) MHD simulation, in face-on (top panels) and edge-on views (bottom) for cores 1 to 6. In contrast to the hydrostatic cores and discs (i.e., n1 and n6) themselves, their low-density surroundings are generally  dominated by magnetic pressure. The outer filamentary structures, which are pressure dominated, are streams with higher densities (see Fig. \ref{fig:final_cores} for the same projections on the same scales).}
    \label{fig:plasma_beta}
\end{figure*}

\begin{figure*}
    \centering
    \includegraphics[width=1.0\linewidth]{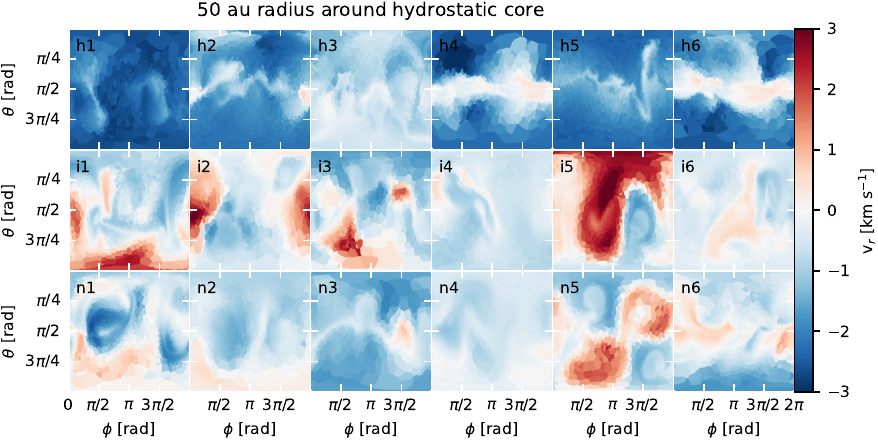}
    \caption{Radial velocity on a 50 au spherical surface around the core in the frame of the first hydrostatic core. $\theta=0,\pi$ correspond to the poles, while $\theta=\frac{\pi}{2}$ is the plane of rotation. In the hydro cases, material moving outwards (red) is, if present at all, part of the disk. In ideal and non-ideal MHD, there are regions of much faster outward motion, although where structures are similar, with reduced speed in the latter case.}
    \label{fig:vrad_shell}
\end{figure*}
Figure~\ref{fig:velocity_profiles} shows azimuthally averaged radial profiles of the measured tangential velocity (in the frame of the first hydrostatic core's angular momentum), and for comparison, the Keplerian velocity (assuming a spherical distribution of all mass, not just the core mass, within the given radius). The simulations without magnetic fields (top panels) usually show rotation almost at the Keplerian speed over tens of au, again except for h3. Also as expected, the rotational speed quickly drops when moving to the outside in ideal MHD, and we find that there is a large variation in non-ideal MHD. We can see that our definition of the disk radius corresponds well to where the tangential velocity starts to drop significantly below the Keplerian one. The two exceptions are n2, where the disk is somewhat asymmetric around its densest region (which can be seen in the projection in Figure~\ref{fig:final_cores}), and n6, where there is an extended region of fast-rotating material, also seen in the corresponding hydro simulation h6. However, this material has very low density (and is thus excluded by our disk definition), as seen in Figure~\ref{fig:density_profiles} in the Appendix.

\subsection{Degree of magnetization}

To quantify the degree of magnetization within the formed hydrostatic cores and disks, we first display the magnetic field strength as a function of radius in Figure~\ref{fig:babs_profiles} for both the ideal and non-ideal MHD simulations. Magnetic field values above 10~G are reached at a radius of 1~au. In some cases, the profile is almost unchanged with the inclusion of ambipolar diffusion. We have checked (in the case of n5, see Appendix~\ref{appendix_checks}) that we restore the typical phenomenon of a plateau lower than 1~G in magnetic field strength in the first hydrostatic core if we remove our limit on the diffusion coefficient, but that at least in our test, this does not actually affect the overall evolution much and there is still no disk that is formed. This is consistent with results from isolated cloud-core collapse \citep{mayer25isolated} that the evolution of the hydrostatic core is not changed significantly by the inclusion of non-ideal MHD in some core geometries (despite a reduction in magnetic field strength). 

In the case of n1 and n6, the magnetic field is suppressed within the region of the inner disk with respect to ideal MHD. This is because while the first hydrostatic core always forms with a relatively high magnetization of 1-10 G, it can then no longer amplify effectively ans is even diffused due to strong diffusion in the core, such that the flux is transported outwards or dissipated. In the central regions of these two simulations ($\lessapprox 5\,{\rm  au}$) `re-coupling' and a subsequent rapid increase in the absolute value of the magnetic field in the inner region are visible at these high densities. Here, high temperatures lead to thermal ionization (which is included in the chemical library we utilize for the coefficients and increases in rate rapidly above $\rho \approx 2 \times 10^{-9}$ g cm$^{-3}$) and raise the ionization fraction to such a degree that ambipolar diffusion becomes negligible. When the magnetic field strength is not markedly reduced by the diffusion, it follows a $\vert B \vert \propto R^{-1.5}$ to  $\vert B \vert \propto R^{-2}$, with the latter expected from non-turbulent collapse with flux-freezing. 

\begin{figure*}
    \centering
    \includegraphics[width=0.7\linewidth]{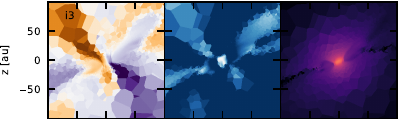}
    \includegraphics[width=0.7\linewidth]{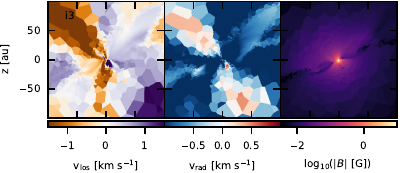}
    \caption{Line-of-sight velocity (left), radial velocity (middle), and absolute value of the magnetic field (right) of the outflow formed in simulation i3 before the launching of the outflow, approximately at the time of formation of the first hydrostatic core (top). The frame is rotated approximately edge-on such that the line-of-sight velocity is a good proxy for the rotation speed. Close to second core formation (bottom), the outflow shows up at large positive radial velocities (bottom middle panel). The localization of outwards-moving material within regions of high rotation speed and perpendicular to the plane of lower magnetic field strength is consistent with a magneto-rotational outflow.}
    \label{fig:outflow_3}
\end{figure*}

The importance of thermal support as compared to the magnetic support is frequently assessed via the dimensionless plasma beta parameter:
\begin{equation}
    \beta = P_{\rm therm} / P_{\rm mag}\,.
\label{eq:plasma_beta}
\end{equation}
Recall that the pressure is a function of density only because we use a barotropic equation of state, given in Equation \eqref{eq: eos}. Figure \ref{fig:plasma_beta} shows $\beta$ for the magnetized runs, both in the plane of the disk and edge-on. The first hydrostatic cores are, even in ideal MHD, dominated by thermal pressure. The same is the case for the parts of the disk found with ambipolar diffusion in regions 1 and 6, reaching $\beta > 10^3$ in the innermost region, although the outer regions of the large disk in region 6 are more magnetically dominated. The low-density surroundings of the hydrostatic core above and below the disk plane, in contrast, show magnetic domination at $\beta < 10^{-3}$. While they are likely magnetically dominated in reality as well, such low values for $\beta$ are a typical feature of simulations, like ours, which do not include radiative transfer \citep[compare e.g.,][]{tu2024fragmentation} as opposed to those which can heat up the surroundings of the disk above the value set by the barotropic equation of state and thereby increase the gas pressure \citep[such as][]{wurster2018collapse}.

\subsection{Outflows}
\label{subsection:outflows}
Magneto-rotational outflows from the first hydrostatic core are a typical phenomenon seen in simulations of protostar formation and have a large effect on the early evolution of protostellar disks as they influence the mass- and angular momentum-budget of the protostar-disk system \citep[][and references therein]{wurster_li2018magnetic}. In isolated cloud core collapse simulations, they have been shown to be strongly influenced by the initial relative orientation of the magnetic field and the rotation axis  \citep{hennebelle2009misalignment,mayer25isolated} and more generally by the presence of initial turbulence \citep{sadanari2023firststars}. In general, departing from the simple geometry of initially aligned axes in a uniformly rotating spherical core is thought to suppress outflows, which should also apply to our significantly more complex prestellar cores.

We have noted above that the trend of an increasing mass in the first hydrostatic core in conjunction with a decline of its specific angular momentum is consistent with the removal of angular momentum by outflows. To confirm the presence of outflows, we map the radial velocity in the surroundings of the hydrostatic core by displaying the radial velocity on a shell of radius 50 au in Figure \ref{fig:vrad_shell} (in the frame of the first hydrostatic core). It is apparent that the hydro simulations show overall greater inflow speeds (consistent with higher net inflow rates, see Figure \ref{fig:inflow} for net inflow on a larger scale), and the only material with a positive radial velocity appears to be within the disk (close to $\theta = \frac{\pi}{2}$) and moving outwards due to its high angular momentum. The picture is very different in the magnetized simulations, where there is gas moving away from the core with speeds of sometimes more than $3\,{\rm km\, s}^{-1}$. While this could be due to relative motions of the hydrostatic core and surroundings, this would lead to a prominent dipole in the figure, which is only clearly seen in the case of i5. As such, the observed distribution of outflowing material on the 50 au shell is consistent with it actually being flung outwards away from the core, likely due to magneto-rotational outflows, as we discuss now. 

In Figure~\ref{fig:outflow_3}, we show the velocity both tangentially (line-of-sight) and radially with respect to the hydrostatic core in the left and middle panels, respectively (upper and lower panels are turned into the same frame, i.e.,~the angular momentum frame of the upper panels). The absolute magnetic field strength on the right panels shows a typical `hourglass' morphology in the inner $\sim 25$ au. Magnetic braking has caused angular momentum to be transported above and below the central plane, giving this material the fastest rotation speed already in the upper panels. There, material is generally still inflowing, though the pressure-supported first core has formed, and there is a plane with slower infall and a relatively low magnetic field strength \citep[a `pseudo-disk', e.g.,][]{xu2021formationI}. This fast-rotating gas starts to escape due to its high speed and can be seen to be moving radially away from the core in the lower panels, in a region generally perpendicular to the pseudo-disk, though somewhat inclined. This is a very typical picture of magnetohydrodynamic outflows, observed among others by \citet{wurster2021nonidealimpactsingle} and \citet{mayer25isolated}. Note that this outflowing material can be matched to that seen in the i3 panel of Figure \ref{fig:vrad_shell}. Other regions show similar morphologies (although often more complex, making it, for example, harder to clearly identify the pseudo-disk). We thus conclude that just as in isolated simulations of cloud-core collapse, magnetically driven outflows can effectively remove angular momentum in our simulations.
\begin{figure*}
    \centering
    \includegraphics[width=0.9\linewidth]{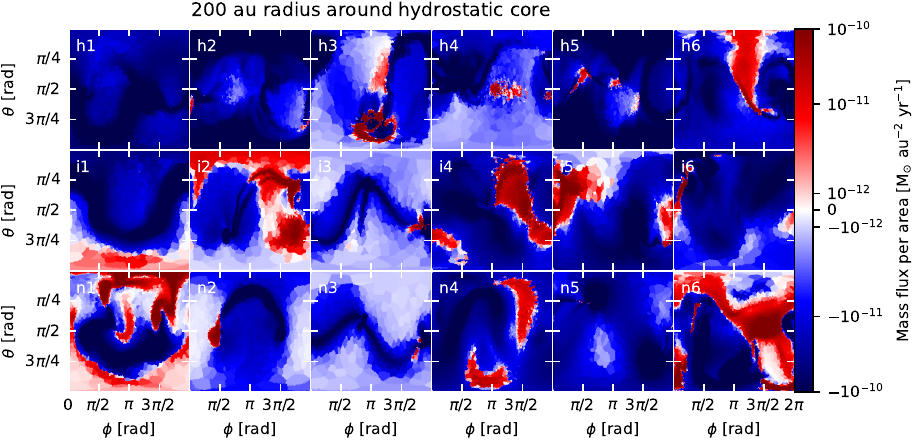}
    \caption{Mass-flux per area on the surface of a sphere with radius 200 au around the first hydrostatic core in rectilinear projections, in the same frame as Fig. \ref{fig:vrad_shell} ($\theta = \frac{\pi}{2}$ in the plane of rotation). Random motions produce negative mass-flux in almost all simulations on this scale, and the accretion is far from uniform.}
    \label{fig:massflux_shell}
\end{figure*}

\subsection{Anisotropic accretion}
\label{subsection:anisotropic_accretion}

The highly non-spherical structure of the initial prestellar core extends down to the scale of the disk. Material is not deposited homogeneously onto the disk in our simulations, but instead is usually mostly accreted from one prominent filament along which the first hydrostatic core forms. This filament then often fragments into smaller fibres. Magnetic fields provide pressure support and thereby suppress this fragmentation, keeping the filament more uniform. In Figure \ref{fig:massflux_shell}, we display the mass-flux per area on a shell of radius 200 AU around the core in the form of rectilinear projections in the final snapshots before second core formation. As with the radial velocities on the 50 au scale that we displayed in Figure \ref{fig:vrad_shell}, and consistent with the integrated values in Figure \ref{fig:inflow}, the mass-flux is generally somewhat lower in the magnetized simulations than in the hydro cases. On this scale, there is some outward mass flux in almost all simulations, likely due to random turbulent motions within the protostellar core. The filament that is prominently visible in some cases in Figure \ref{fig:final_cores} is also seen here (most obviously in i2/i3, and n2/n3) and is a region of increased mass-influx, reflecting its comparatively high density. Overall, the inflow of gas to the forming protostar is clearly not isotropic, and at least some of the structures at this scale connect down to the first hydrostatic core / disk, such as the filament.

\section{Structure of disks in non-ideal MHD}
\label{sec:results_specific}

\subsection{Toomre analysis}

All simulations in which a prominent disk is formed show clear substructure in the form of spiral arms. Such structures can be sites of gravitational fragmentation in self-gravitating disks. The susceptibility of a disk to fragmentation  can be assessed with the Toomre $Q$ parameter \citep{toomre1964q}:
\begin{equation}
    Q = \frac{c_{\rm s} \kappa}{\pi G \Sigma}\,,
    \label{eq:Toomre}
\end{equation}
where $c_{\rm s}$ is the sound speed, which we calculate as the isothermal sound speed $c_{\rm s} \equiv c_T = \sqrt{\frac{P}{\rho}}$ (this, however, does take into account the change in $P$ as a function of $\rho$ given by the equation of state), and $\kappa$ is the epicyclic frequency, which we approximate with the Keplerian speed at a given radius, $\kappa \approx \Omega_{\rm K} =  \sqrt{\frac{G M(<r)}{r^3}}$. To determine the local surface density $\Sigma$, we integrate over a scale height of 5 au, and average $c_{\rm s}$ (weighted by density) over the same vertical range. In the presence of magnetic fields, we take the additional support from magnetic pressure into account by modifying the sound speed  in equation~\eqref{eq:Toomre} to $\tilde{c}_{\rm s} = \sqrt{c_{\rm s}^2 + {\rm v}_{\rm A}^2}$, where  ${\rm v}_{\rm A} = \frac{B}{\sqrt{\mu_0 \rho}}$ is the Alfv{\' e}n speed\footnote{The extra contribution from magnetic pressure can be written as $\frac{Q_{\rm mag}}{Q} = \sqrt{1 + \frac{2}{\beta}}$, where $\beta$ is the plasma beta as defined in equation \eqref{eq:plasma_beta}.}. $Q$ can be thought of as a ratio between forces that prevent local collapse (pressure and rotation in the numerator) and gravitational forces that induce collapse (in the denominator), such that $Q < 1$ indicates a region where fragmentation might occur. 

\begin{figure*}
    \centering
    \includegraphics[width=0.9\linewidth]{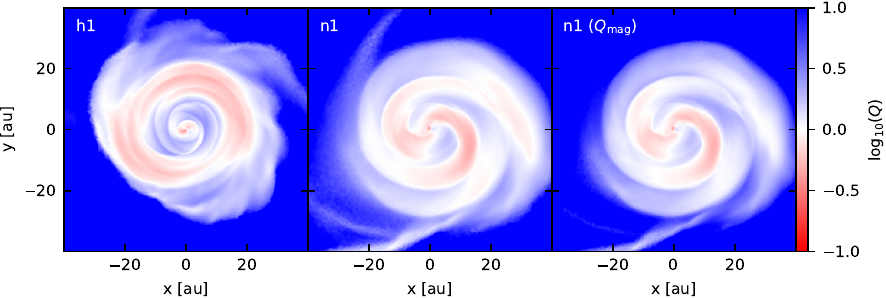}
    \includegraphics[width=0.9\linewidth]{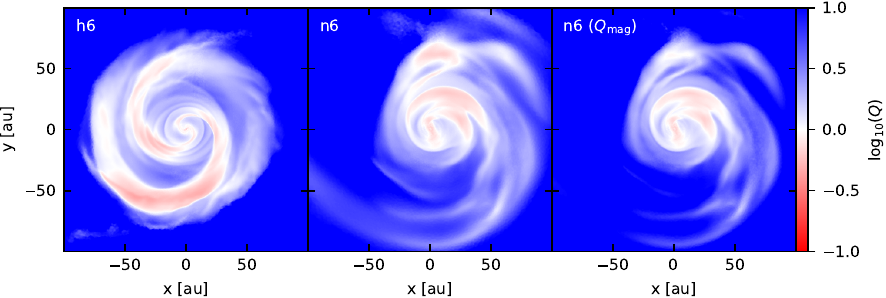}
    \caption{Face-on projection of the Toomre $Q$ parameter of core 1 (top) and core 6 (bottom) with hydrodynamics (left) and non-ideal MHD (middle). The right panels show the magnetic $Q$ including the Alf Alfv{\' e} in the sound speed. Unstable regions with $Q < 1$, indicating that the discs could become subject to fragmentation in the spiral arms.}
    \label{fig:toomre}
\end{figure*}

In Figure \ref{fig:toomre}, we show the hydrodynamical and magnetic Toomre~$Q$ of disks formed in our simulations. We can see that the disks are dense enough in certain areas to become locally gravitationally unstable, and these areas are localized to spiral arms. The outer region of the disk in simulation n6 is more magnetized than in n1, as indicated here by the more substantial increase in $Q$ when the magnetic contribution is included.

\subsection{Interplay between disk and magnetic field}

\begin{figure*}
    \centering
    \includegraphics[width=0.37\linewidth]{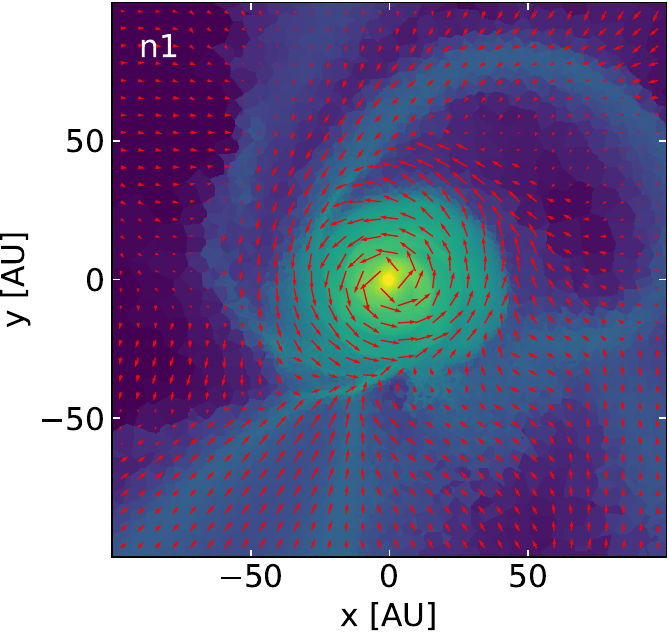}
    \includegraphics[width=0.472\linewidth]{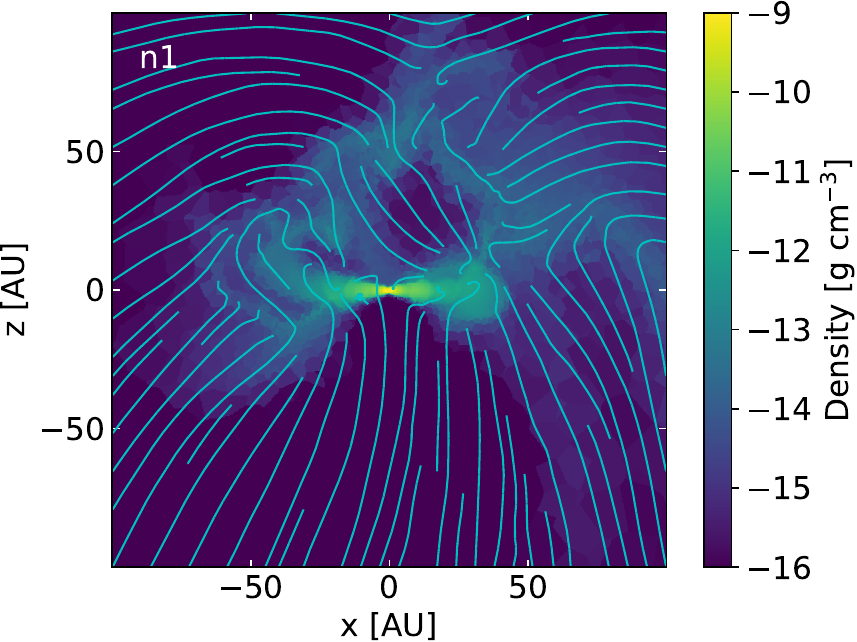}
    \includegraphics[width=0.37\linewidth]{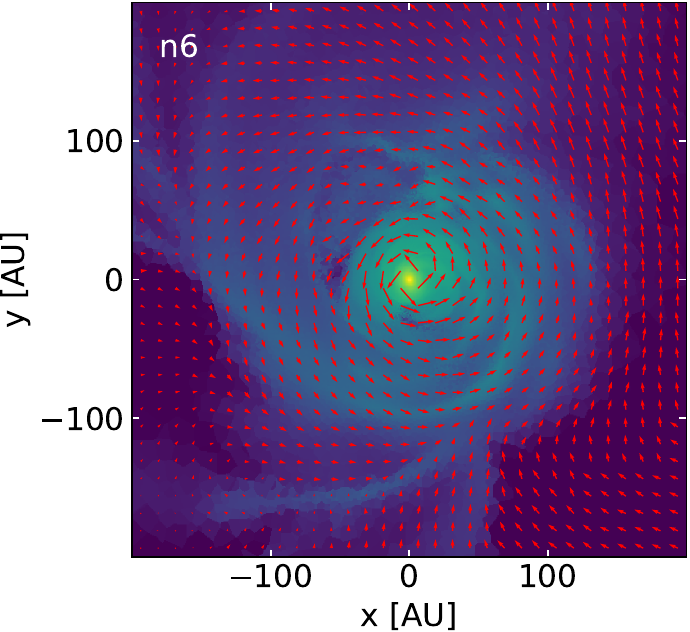}
    \includegraphics[width=0.472\linewidth]{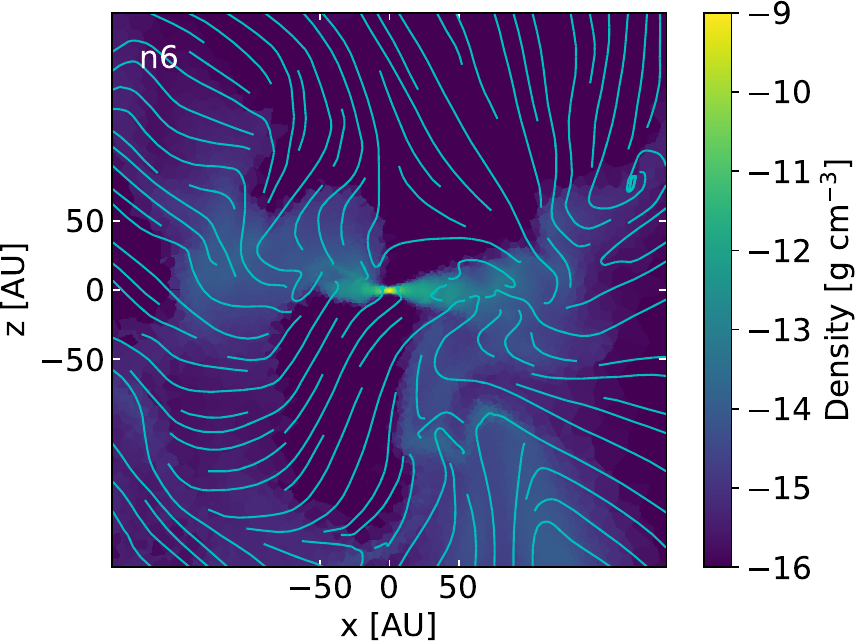}
    \caption{Density slices (face-on on the left, edge-on on the right) of the disk resulting from regions 1 (top) and 6 (bottom) with ambipolar diffusion. Note the different diameters of the upper and lower panels. The in-plane velocity is shown as red arrows in the left panels, while the in-plane component of the magnetic field lines is displayed as cyan lines in the right panels. The tangential velocity component is much higher than the radial one in the disk.}
    \label{fig:planes}
\end{figure*}

In simulations of isolated core collapse, the relative orientation of the magnetic field and angular momentum vector of the initial core has been shown to significantly alter the evolution \citep{hennebelle2009misalignment,mayer25isolated}. Figure~\ref{fig:planes} shows both face-on and edge-on slices of the two largest disks formed with ambipolar diffusion (regions 1 and 6), displaying density, velocity vectors (face-on), and streamlines from the in-plane components of the magnetic field (edge-on). In the face-on views, we can identify a substantial rotational component when compared to the radial velocity component, indicating strong rotational support as expected from the tangential velocity profiles in Figure~\ref{fig:velocity_profiles}. In the edge-on views, we observe that there is no strong pinching of the magnetic field, and the orientation appears to be somewhat more in-plane in the case of n1 than in that of n6. In both cases, we see a dense central structure (the hydrostatic core) surrounded by a disk with lower density. In n6, especially the disk departs quite strongly from axisymmetry, and there are dense filaments in the midplane connecting to the disk.

\begin{figure*}
    \centering
    \includegraphics[width=0.45\linewidth]{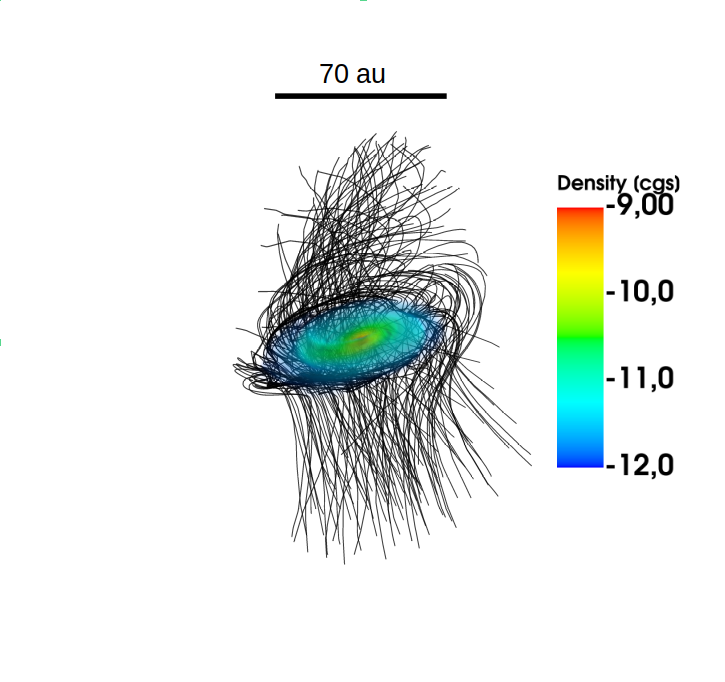}
    \includegraphics[width=0.45\linewidth]{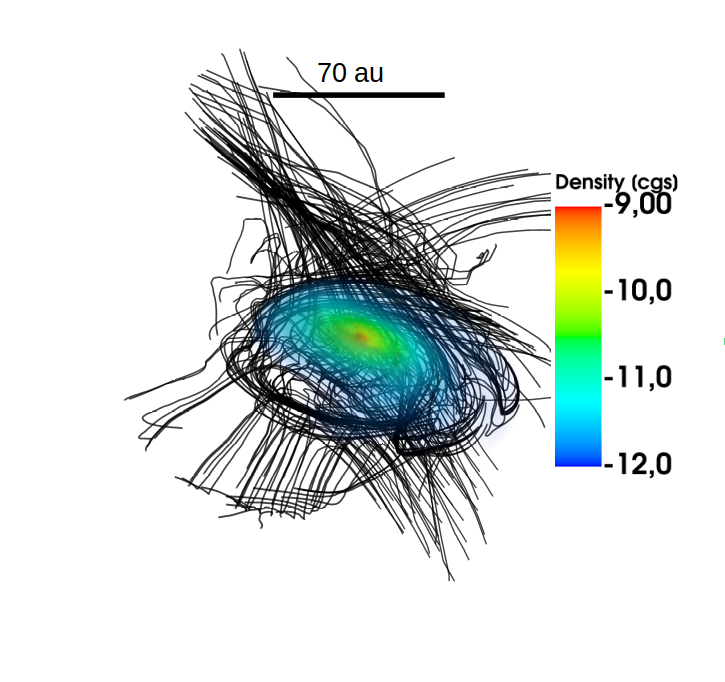}
    \caption{Density distribution and magnetic field lines of the non-ideal MHD disks in regions 1 and 6. Shown 16$\times$16 field lines that intersect a spherical surface of radius 20 au. In both cases, the magnetic field near the disk is largely toroidal, but the imprint of the large-scale field is also still visible. The viewing angle is chosen for best visibility.}
    \label{fig:streamplots}
\end{figure*}

To obtain a full picture of the very complex magnetic field structure that is formed as a result of the combination of turbulent infall and rotation, we show a 3D view of the disks formed in the runs with ambipolar diffusion in regions 1 and 6 in Figure \ref{fig:streamplots}. We can see that the rotation has twisted the magnetic field in and near the disk substantially. In the case of n1, the twisting is more pronounced on one side of the disk than the other, which also appears to be the direction into which it launches its outflow (which also has substantial rotation). In the case of n6, we observed and elongated structure of relatively uniform field direction, which is likely connected to a density feature discussed in the next subsection.

\subsection{Example of streamer-like morphology}

Anisotropic accretion of material along narrow filaments onto protostellar accretion disks is by now a well-observed phenomenon (see references in the introduction). Simulations can be set up to specifically model such a scenario \citep[e.g.,][]{grigoryev2024streamsimulation} which is observed in high-resolution simulations of longer-term disk evolution \citep{heigl2024disk}. But if filaments form close to the birth of the protostar, along with the disk itself, they should also occur naturally in our simulations. 

\begin{figure}
    \centering
    \includegraphics[width=1.0\linewidth]{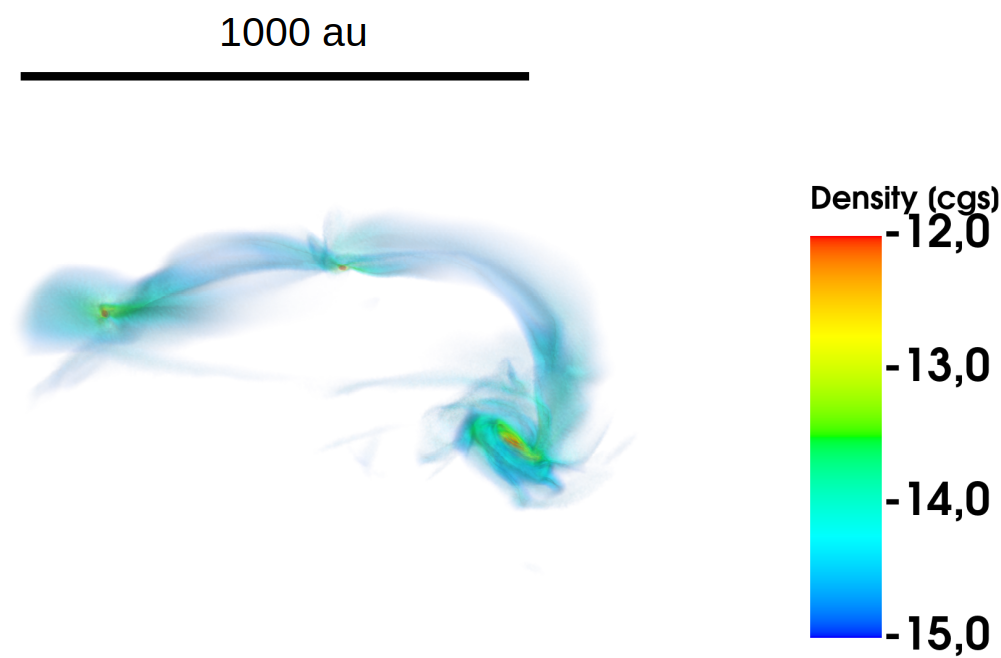}
    \caption{A 3D volume rendering showing all material with $\rho > 10^{-15}\,{\rm g\, cm}^{-3}$ up to approximately 1000~au from the centre of the disk in the non-ideal MHD simulation n6 (on the bottom right). There is one large structure connecting to the disk (the `streamer') as well as multiple smaller ones. We also clearly observe that the initial core region has fragmented into several forming protostars, which are situated along the `streamer'. The viewing angle is chosen for best visibility.}
    \label{fig:streamer_6}
\end{figure}

\begin{figure}
    \centering
    \includegraphics[width=1.0\linewidth]{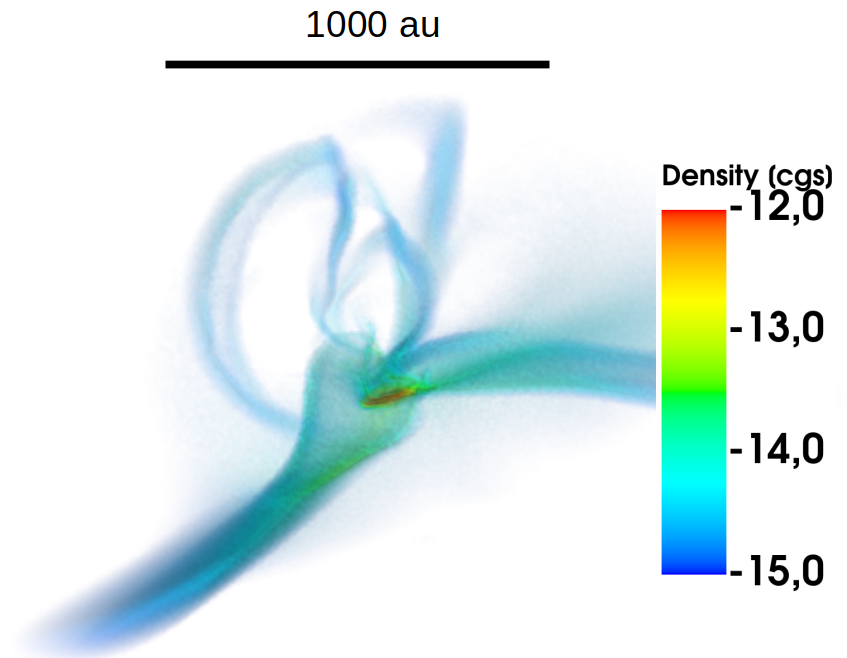}
    \caption{A 3D volume rendering showing all material above a density of $10^{-15}\,{\rm g\, cm}^{-3}$ up to approximately 1000~au from the centre of the disc in the non-ideal MHD simulation n1 (centre). In this case, the disk is connected to two large filaments, as well as smaller streams, originating from the disk (visible in the top middle). The viewing angle is chosen for best visibility.}
    \label{fig:streamer_1}
\end{figure}

In Figure \ref{fig:streamer_6}, we show a volume rendering of the density  distribution from the final snapshot of simulation n6, which shows an elongated structure connected to the large disk. Along this filament, two other hydrostatic cores have formed. This morphology is similar to `bridges' seen in the simulations of multiple formation by \cite{kuffmeier2019bridge} \citep[illustrated in a similar rendering to ours in Fig.~19 of][]{pineda2023bubbles} and observed by \cite{gieser2024streamer}. The infall-velocity of this structure onto the disk is in the range of 1-2 ${\rm km \, s}^{-1}$. Figure \ref{fig:streamer_1} shows a similar visualization, but of simulation n1, in which case the disk is embedded within a filament and is, as such, connected to a dense structure on two sides. We also find smaller streams of material which are likely associated with the outflow or interchange instabilities \citep[see e.g.][]{machida2025interchange,vaytet2018protostellar} as they originate from the disk and travel outwards (as found from examining earlier snapshots). We plan to turn this qualitative examination of the occurrence of streamer-type structures into synthetic observations in the future.

\subsection{Comparison between magnetic field models on the disk and stellar core scale}

\begin{figure*}
    \centering
    \includegraphics[width=0.8\linewidth]{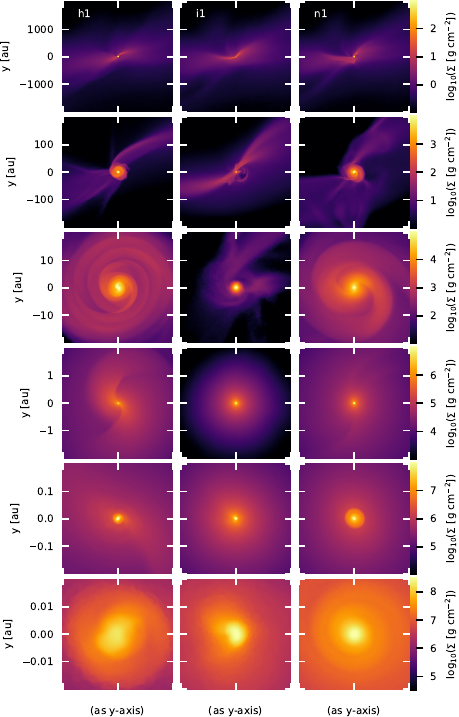}
    \caption{Visual zoom sequence  of the final cores of region 1 for the hydrodynamic (h, left), ideal-MHD (i, middle), and non-ideal MHD (n, right) simulations. The systems are shown from scales of the protostellar ($\sim 10^3$ au, top) down to the stellar core ($\sim 10^{-2}$ au, bottom). Differences are apparent on all of the depicted scales. There is no large disk in ideal MHD, while hydrodynamics and non-ideal MHD both have disks which are large, but mutually distinct in structure. These differences are present up to the scale of the stellar core.}
    \label{fig:moneyplot}
\end{figure*}

To show how the second core connects to its surrounding disk and the larger-scale envelope, we show column density projections at scales from 4000 au in diameter down to 0.04 au ($< 10\, {\rm R}_\odot$) for core 1 in Figure~\ref{fig:moneyplot} (note that different amounts of time have elapsed from the beginning of the zoom-in between these setups and these are from slightly later snapshots than shown previously). We have run all simulations except for n6 until this point, but choose to show core 1 as the most interesting example. In the case of hydrodynamics and non-ideal MHD, in addition to the disk surrounding the first hydrostatic core at the scale of tens of au, there is a much smaller (< 0.1 au) disk surrounding the second core, sometimes called the `second disk'. This simultaneous occurrence of both of these disks has been observed before and referred to as `nested disks' \citep{bhandare2024mixing}. It also visually resembles the structures observed in the non-ideal MHD simulations of \cite{ahmad2025birth}, although in our case this is seen at a much lower resolution, which indicates that the size of the protostellar core is likely not correctly estimated, not least because we do not include radiation \citep{ahmad2023protostar}. Nevertheless, we conclude that stellar cores with the approximate size of a solar radius ($R_\odot = 0.05$~au), as found by \citet{tomida2013rmhd,vaytet2018protostellar,wurster2018originI,ahmad2023protostar} and others, are also formed in our simulations. This is likely a consequence of the fundamental physics underlying the formation of the second core (the hardening of the equation when the dissociation of H$_2$ is complete), in our case set by the barotropic equation of state.

\section{Limitations and future work}
\label{sec:discussion}
The first obvious limitation of our study is our lack of a statistically significant sample, which is a consequence of the high resolution and thus computational cost of our simulations, corroborated by the fact that we actually resolve the central region of the disk without using sink particles. We have thus focused on a relatively limited number of objects, precluding us from drawing very general conclusions from this sample. Among the 6 cases we examined, there is already a large diversity in evolution in terms of disk sizes, structure, and even whether cores exist at all. Entirely distinct protostar formation scenarios are therefore definitely possible, and we expect a rich variety in the specific characteristics of individual simulated and observed objects.

The turbulent driving we utilized in this work is very simple and only gives a rough approximation of actual feedback processes in the ISM. However, we consider it already an improvement over the often-used method of just including a turbulent velocity field in the initial conditions of an isolated cloud core collapse. Likewise, it should be more realistic than driving turbulence via an ad-hoc Ornstein-Uhlenbeck process \citep[as used by e.g.,][]{yang2025zoom}. An important limitation of our approach is that supernovae do not actually happen randomly throughout the ISM, but are associated with star formation. It has been shown that choosing the peaks of the density distribution as injection sites also does not give the most realistic ISM, however, which is best modelled with a mixed driving that also models the effect of runaway stars \citep{gatto2015ism}. We plan to include this more complex feedback localization in the future; however, an alternative and more realistic approach would be to directly include actual star particles formed in the simulation (in contrast to our current sink particles, which do not inject any energy). For our simulation to more closely model a section of the ISM in a galaxy, a different boundary condition and gravitational potential should be used more akin to the tall-box setups of e.g. \cite{walch2015silcc,kim_ostriker2017tigress}. In our model, gas cannot escape in what would be the vertical direction in a galactic disk, and the crucial importance of the vertical pressure-budget is not modelled.

We do not include radiative transfer in our simulations, and as such, our disks are likely colder than expected \citep{whitehouse2006thermodynamics}. This, in turn, reduces their pressure support (lowering their Toomre $Q$), and more easily leads to GI and spiral structure even at low surface densities, but is expected to also generally influence the evolution of the disk \citep[][]{bate2014collapse}.

In this work, we have focused on the influence of the magnetic field treatment within prestellar cores. To obtain a proper and fully consistent comparison between an ISM with hydrodynamics, ideal, and non-ideal MHD, one would also need to run the driving and tracing of the zoom-in candidates with the different magnetic field models. While somewhat slower, we have tested that running the early evolution with non-ideal MHD is possible without numerical issues. We aim to explore the ISM structure under the influence of ambipolar diffusion in the future, but have here decided to focus on differences that are a result of non-ideal MHD on the scale of the protostellar cores, and, as such, needed identical initial conditions for the zoom-in.

A natural generalization of our model is to include Ohmic diffusion and the Hall effect. The influence of the latter heavily depends on the particular configuration of a prestellar core, and we plan to carry out simulations exploring this in the future, similar to what has already been done in the isolated simulations by \cite{mayer25isolated}. It was also shown therein, as well as in previous work \citep[e.g.,][]{wurster2018ionizationrates,zhao2021interplay}, that the impact of non-ideal MHD depends heavily on properties of the grain size distribution, gas microphysics details, and the cosmic ray ionization rate, which are fraught with uncertainty and, in the case of the latter, certainly differ from core to core. \cite{nishio2025cosmicrays} have recently shown that modelling a non-constant cosmic-ray ionization rate on the protostellar core and disk scale leads to significant differences from assuming a constant rate.  As such, our model with ambipolar diffusion should be seen as only one representative scenario, while the hydrodynamical and ideal MHD simulations bracket the range of possible behaviour. In addition to that, there are other physical processes that play a role in the model evolution and which we do not consider yet, such as the effects of non-equilibrium chemistry in the cooling and how this interacts with the detailed dust physics.

Finally, because we do not use sink particles in the zoom-ins, it is not feasible to follow the evolution of the emerging protostar beyond the moment of its birth. We therefore cannot make any statements about the subsequent evolution of its accretion disk or lack thereof. Studies that followed the evolution in ideal MHD for a longer amount of time \citep[e.g. ][]{seifried2013turbulence,kueffmeier2017zoom,yang2025zoom} found disks forming in ideal MHD, although it is apparent from Figure 2 of the latter that it does not form early on. We plan to extend our study to the further evolution of the disks in the future. On the other hand, not using sink particles allows us to avoid some of their numerical pitfalls \citep[see, e.g. ][]{machida2014sinks} and to actually resolve the inner disk.

\section{Conclusions}

\label{sec:conclusions}
We have run simulations that self-consistently model the collapse to protostars starting from a supernova-driven turbulent ISM, and analyzed the structure of the resulting hydrostatic cores and, where they form, their surrounding disks. Our main conclusions are as follows:
\begin{enumerate}
\item All simulations without magnetic fields have formed disks at the moment of stellar core formation. However, the disks vary greatly in size and morphology, from disks of a few au with strongly wound-up spiral arms up to disks more than 100 au in size with a `grand-design' appearance.
\item Outflows are a general feature in simulations including magnetic fields, being launched around the time of first core formation. They can propagate for hundreds of AU in the time to second core formation. Outflow speeds are lower for the runs with ambipolar diffusion, but the outflows are not fully suppressed. Where they happen, they can completely shut off disk formation by carrying away sufficient amounts of angular momentum or, in other cases, delay the collapse to the second core significantly by removing mass.
\item Non-ideal MHD does, in some, but not all, cases, lead to the formation of a disk where the corresponding ideal MHD simulation does not. If disks form with ambipolar diffusion, they are of similar size to the corresponding simulation without magnetic fields. The structure of these disks, in particular the spiral substructure, is clearly distinct between cases with and without magnetic fields, and as such, non-ideal MHD forms a markedly separate regime from both hydrodynamics and ideal MHD in the turbulent environments we studied.
\item The global characteristics of the initial cores are of limited utility for predicting the occurrence of disks at this early phase of the evolution. Instead, the individual geometry of a core, including whether it forms at a shock front, appears to be more relevant in setting how much mass actually reaches the (possible) disk before stellar core formation.
\end{enumerate}

Our simulations indicate that if ideal MHD were a good description of reality, stars would never be been born with a disk -- even though they might still form later, which is beyond the scope of our study. This implies that the suppression of disk formation in ideal MHD is not limited to the idealized setups that have occasionally been identified as the main cause of the `magnetic braking catastrophe', but that for realistic levels of magnetization, the formation is at least heavily suppressed and delayed in any geometry or possibly completely prevented. As such, this underlines the central importance of non-ideal MHD in these turbulent environments for allowing an early formation of large protoplanetary disks. 

\section*{Acknowledgements}
The authors acknowledge helpful discussions with Troels Haugbølle and Michael Küffmeier. TN and PC acknowledge the support of the Deutsche Forschungsgemeinschaft (DFG, German Research Foundation) under Germany’s Excellence Strategy - EXC-2094 - 390783311 of the DFG Cluster of Excellence ''ORIGINS''. 
Support for OZ was provided by Harvard University through
the Institute for Theory and Computation Fellowship.
SW gratefully acknowledges support by the CRC 1601 (SFB 1601 sub-project A5) funded by the DFG (German Research Foundation) – grant number 500700252.
\section*{Data Availability}
The data underlying this paper will be shared upon reasonable request to the corresponding author.

\bibliographystyle{mnras}
\bibliography{main.bib}

\appendix

\section{State of the ISM after the end of driving}
\label{appendix_after_drving}

\begin{figure}
    \centering
    \includegraphics[width=0.95\linewidth]{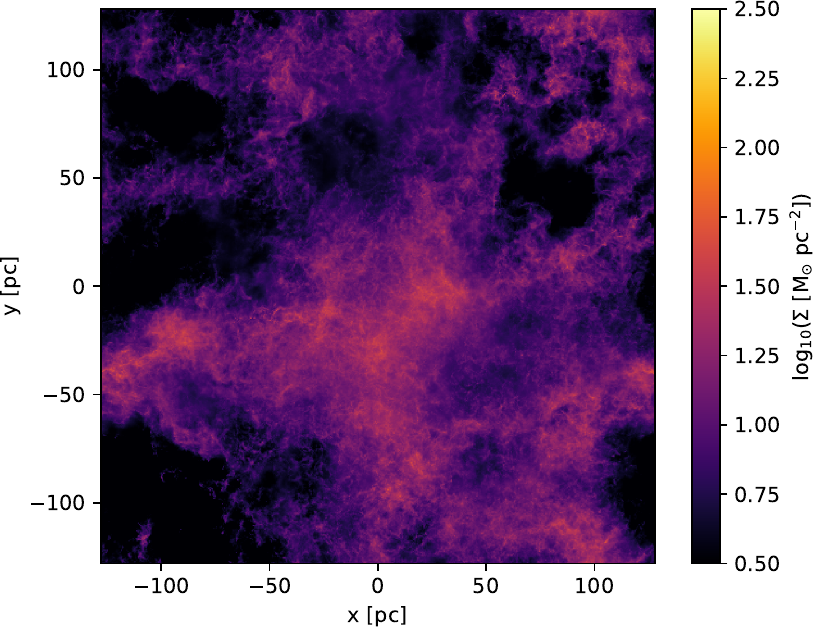}
    \includegraphics[width=0.95\linewidth]{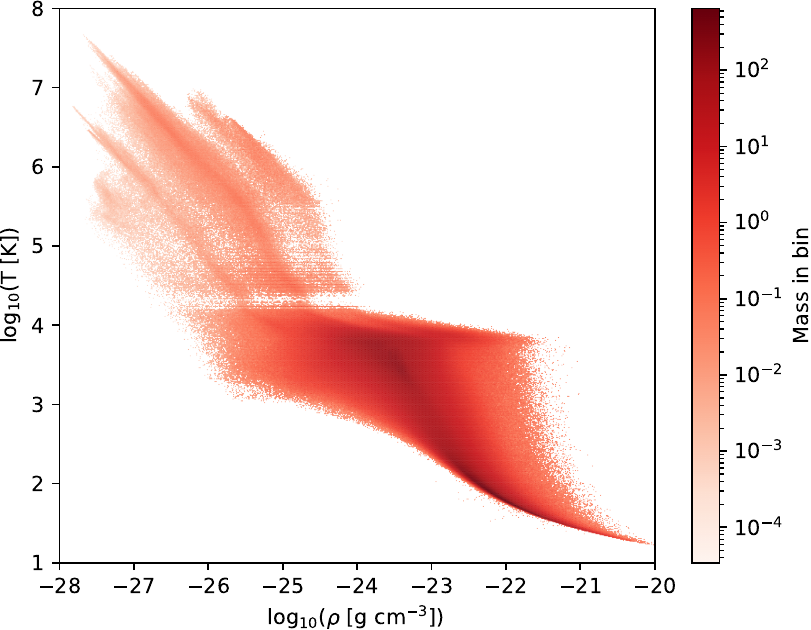}
    \caption{Structure of our ISM after the end of the driving. The top panel shows the column density of the entire box, while the bottom panel shows the density-temperature phase diagram.}
    \label{fig:after_driving}
\end{figure}

To showcase the ISM we obtain from our driving procedure and cooling prescription, we display column density projections and the density-temperature phase-space in Figure \ref{fig:after_driving}. At this point, there is still a hot phase from recent supernovae, which has cooled down until the first sink forms. A comparison with the column density of the entire box in Figure \ref{fig:bigplot} shows that there is still significantly less structure, which is later formed under the influence of gravity (which is only turned on starting at this point).

\section{Radial density profiles of disks}
\label{appendix_radial_dens}
\begin{figure*}
    \centering
    \includegraphics[width=1.0\linewidth]{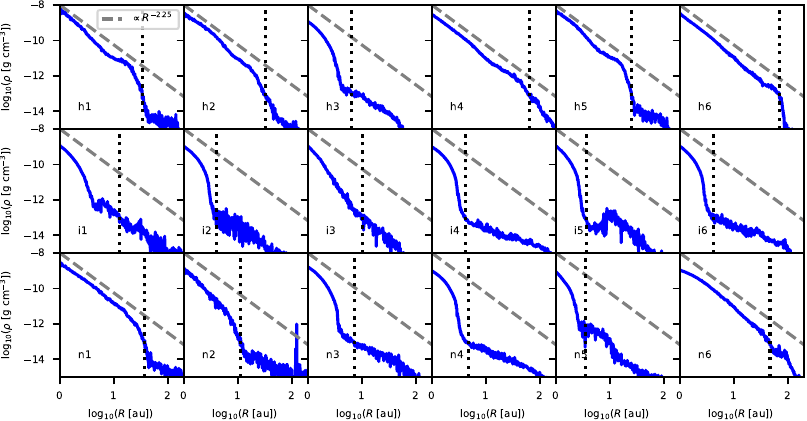}
    \caption{Density as a function of radius. Dotted vertical lines indicate the disk size.}
    \label{fig:density_profiles}
\end{figure*}

Figure \ref{fig:density_profiles} shows radial density profiles of all disks / hydrostatic cores at the end of the simulation. Within the disk radius, we roughly see a scaling of $\rho \propto R^{-2}$ to  $\rho \propto R^{-2.5}$. In most cases, the disk edge is associated with a sharp drop in density along with a drop in rotation speed, as discussed in the main part of the paper, see Figure \ref{fig:velocity_profiles}.

\section{Influence of the choice of starting point for zoom-in, resolution, and coefficient-limit}
\label{appendix_checks}

\begin{figure*}
    \centering
    \includegraphics[width=0.4\linewidth]{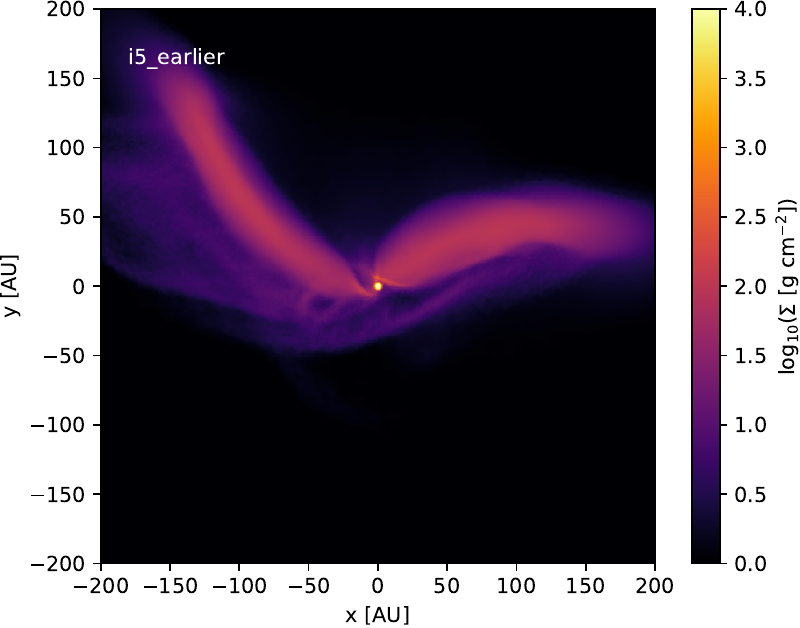}
    \includegraphics[width=0.4\linewidth]{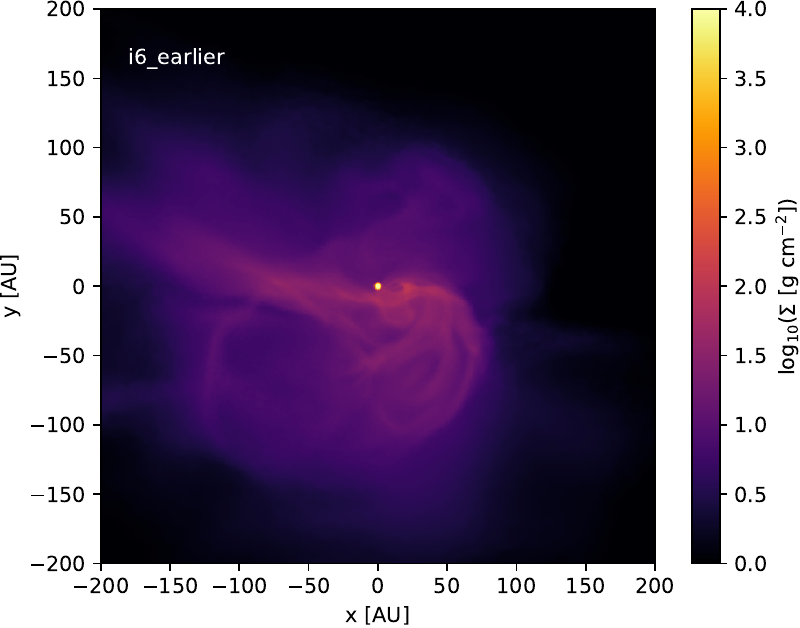}
    \includegraphics[width=0.4\linewidth]{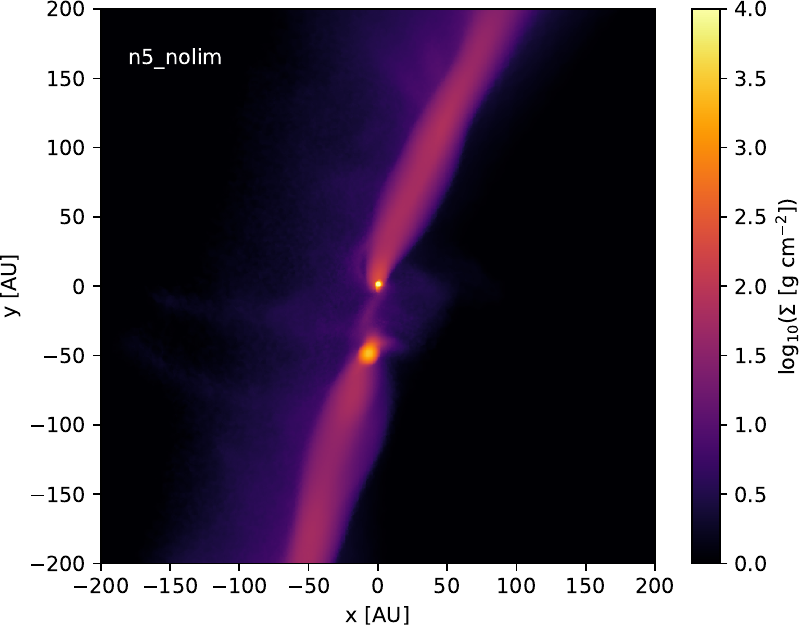}
    \includegraphics[width=0.4\linewidth]{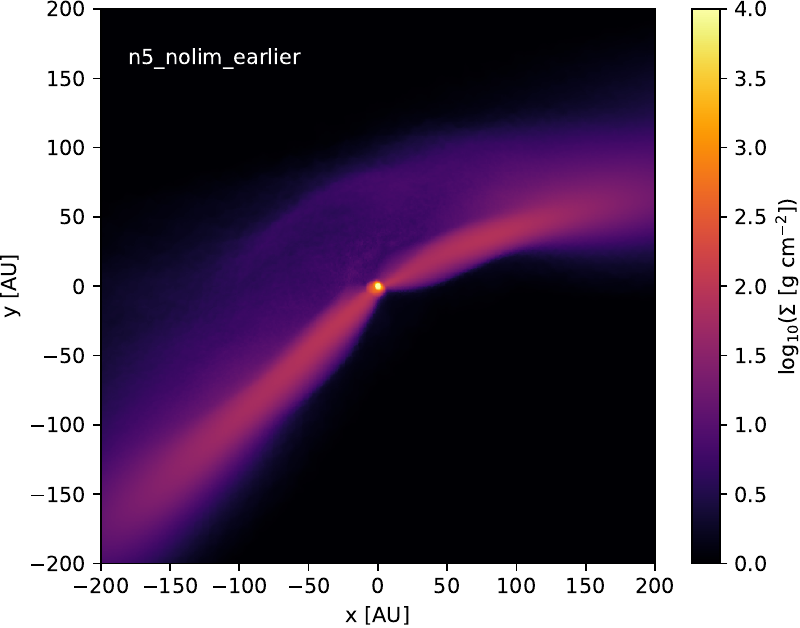}
    \caption{Column density projections in our additional tests. Upper left and right: Repeat of simulations i5 and i6, respectively, where the zoom-in is started (and therefore the volume limit removed) approximately 20 kyr earlier. Bottom Left: Repeat of simulation n6 with no limit on the ambipolar diffusion coefficient and the additional inclusion of Ohmic diffusion. Bottom right: Another rerun of simulation n6 without a limit, but also started  approximately 20 kyr earlier.}
    \label{fig:tests}
\end{figure*}

In Figure~\ref{fig:tests} we show projections similar to those in Figure \ref{fig:final_cores} (only face-on here) for four additional tests. On the middle and right, we started simulations i5 and i6 significantly earlier than in the body of the paper (both on the order of tens of kyr before sink formation), which also means that the resolution is increased in the collapse of the core, as the volume limit is never applied. The resulting general morphology is quite similar, in particular there is still a prominent filament in the middle as in i6, and an elongated stream of material on the right as in i7 (which likely originates from the same large-scale structure as the `streamer' we display in Figure~\ref{fig:streamer_6}). While it would be preferable to start all our simulations before the application of the volume limit, this would make the hydro run essentially not comparable at all to the magnetized runs (as the protostellar cores would already be very different), and running the box without any volume limit is currently not feasible.

In the bottom panels of the Figure, we removed the limit on the ambipolar diffusion coefficient from the physical value and even added Ohmic diffusion. The bottom left is started from the same point as the simulation in the body of the paper, while on the bottom right, we start from an earlier snapshot, as for the top left panel. We have checked that the magnetic field within the core reaches the typical maximum at less than 1~G and is kept from further increase until the recoupling, but there is still no discernible disk in both cases. The second hydrostatic core that is forming along the filament on the bottom left is likely a result of a slightly delayed collapse time, as we also find this behaviour in i5, which does indeed also take somewhat longer to collapse than n5 \citep[we showed in][that non-ideal MHD simulations may collapse both faster and slower than the same setup in ideal MHD]{mayer25isolated}. In the bottom right, this second core has just merged with the first one, contributing to triggering the second collapse in this case.

In conclusion, while both the use of a limit on the diffusion coefficient and the late starting point / reduced resolution affect details of the resulting cores, they do not change our general conclusions about the lack of disks in ideal MHD, or the fact that non-ideal MHD does not guarantee the presence of a disk either.

\label{lastpage}
\end{document}